\documentclass[%
reprint,
superscriptaddress,
 amsmath,amssymb,
 prl,
]{revtex4-2}

\usepackage{graphicx}
\usepackage{dcolumn}

\usepackage{color}
\usepackage{bm}
\usepackage{amssymb}
\usepackage{amsmath}

\newcommand{\de}{\partial}
\newcommand{\divg}{\bm{\nabla} \cdot} 
\newcommand{\curl}{ \bm{\nabla} \times}
\newcommand{\grad}{\bm{\nabla}}
\newcommand{\g}{\textbf}

\newcommand{\jgr}{Journal of Geophysical Research}

\renewcommand{\aa}{\bm{a} }
\newcommand{\bb}{\bm{b} }

\newcommand{\ff}{\bm{f} }
\renewcommand{\gg}{\bm{g} }
\newcommand{\jj}{\bm{j} }

\newcommand{\vv}{\bm{v} }

\newcommand{\zz}{\bm{z} }
\newcommand{\bomega}{\bm{\omega} }
\newcommand{\bOmega}{\bm{\Omega} }

\usepackage[mathlines]{lineno}

\begin{document}

\preprint{XXX}

\title{Relaxation of the turbulent magnetosheath}

\author{Francesco Pecora}
\affiliation{Department of Physics and Astronomy, University of Delaware, Newark, DE 19716, USA}
\email{fpecora@udel.edu}

\author{Yan Yang}
\affiliation{Department of Physics and Astronomy, University of Delaware, Newark, DE 19716, USA}

\author{Alexandros Chasapis}
\affiliation{University of Colorado Boulder, Boulder, CO 80309, USA}

\author{Sergio Servidio}
\affiliation{Università della Calabria, Arcavacata di Rende, 87036, IT}

\author{Manuel Cuesta}
\affiliation{Department of Physics and Astronomy, University of Delaware, Newark, DE 19716, USA}

\author{Sohom Roy}
\affiliation{Department of Physics and Astronomy, University of Delaware, Newark, DE 19716, USA}

\author{Rohit Chhiber}
\affiliation{Department of Physics and Astronomy, University of Delaware, Newark, DE 19716, USA}
\affiliation{NASA Goddard Space Flight Center, Greenbelt, Maryland 20771, USA}

\author{Riddhi Bandyopadhyay}
\affiliation{Department of Astrophysical Sciences, Princeton University, Princeton, NJ 08544, USA}

\author{D.~J. Gershman}
\affiliation{NASA Goddard Space Flight Center, Greenbelt, Maryland 20771, USA}

\author{B.~L. Giles}
\affiliation{NASA Goddard Space Flight Center, Greenbelt, Maryland 20771, USA}

\author{J.~L. Burch}
\affiliation{Southwest Research Institute, San Antonio, Texas 78238, USA}

\author{William H. Matthaeus}
\affiliation{Department of Physics and Astronomy, University of Delaware, Newark, DE 19716, USA}

\date{\today}

\begin{abstract}

In turbulence, nonlinear terms drive energy transfer from large-scale eddies into small scales through the so-called energy cascade. Turbulence often relaxes toward states that minimize energy; typically these states are considered globally. However, turbulence can also relax toward local quasi-equilibrium states, creating patches or cells where the magnitude of nonlinearity
is reduced and energy cascade is impaired. We show, for the first time, compelling observational evidence that this ``cellularization'' of turbulence can occur due to 
local relaxation in a strongly turbulent natural environment such as the Earth's magnetosheath.

\end{abstract}

\maketitle


\textit{Introduction.--}Turbulence is  disordered, but may also involve relaxation processes that drive the system toward statistically predictable states \citep{matthaeus2012review}. The problem of turbulent relaxation has been addressed from theoretical \citep{frisch1975possibility, montgomery1978three, stribling1991statistical} and numerical \citep{matthaeus1980selective, ting1986turbulent, stribling1991relaxation} perspectives. Relaxation may lead to the emergence of long-time metastable quasi-equilibrium states. However, there is evidence that \citep{PelzEA85, MontEA92-pof, servidio2008depression, MattEA08-align, ServidioEA10-PF} partially relaxed states can emerge on time scales 
on the order of a large-scale nonlinear (eddy turnover) times. Rapid local relaxation principles can contribute to the understanding of nonlinear processes in space plasma, astrophysical and geophysical systems, by accounting for the generation of distinctive measurable correlations. Here we present observational evidence that several forms of rapid relaxation -- including  Alfv\'enic, Beltrami, and force-free states -- occur in the highly turbulent terrestrial magnetosheath plasma.

A canonical example 
is two-dimensional (2D) hydrodynamics, which exhibits both slow emergence of metastable global maximum entropy states \citep{MontJoyce74, MontEA93} as well as a local multiscale rapid relaxation \citep{ServidioEA10-2Dns}. This system is relevant to laboratory plasmas \citep{RodgersEA09} geophysical flows, ionospheric structure \cite{KintnerKelley85}, and other self-organizing systems \citep{BouchetVenaille12}.

Two classes of relaxation have been discussed in magnetoydrodynamics (MHD): (I) Selective decay processes \citep{montgomery1978three, MattMont80} in which a global long-wavelength relaxed state emerges by minimization of one ideal invariant (e.g., energy) relative to another (e.g., magnetic helicity) \citep{chandrasekhar1958force}; an example is Taylor relaxation \cite{taylor1974relaxation, Taylor86}; and (II) dynamic alignment processes, in which a distinctive correlation, such as large amplitude Alfv\'enic fluctuations, emerge dynamically \citep{woltjer1958hydromagnetic, PouquetEA86}. Both types reduce the 
amplitude of nonlinearities \citep{stribling1991relaxation}. 

Plasma relaxation is also relevant to solar and interplanetary studies,
where
for example, 
magnetic clouds are typically modeled as force-free Lundquist states \cite{burlaga1981magnetic}. 
Taylor relaxation may 
drive these somewhat isolated magnetic flux tubes toward force-free states \cite{HeyvaertsPriest84}. A broader class of target Grad-Shafranov equilibria 
are force-balance states, 
\cite{grad1958hydromagnetic}, which can be identified 
in solar wind data \cite{sonnerup2016reconstruction, HasegawaEA14}. Maximum entropy states \cite{MontEA79} are also purported to emerge through turbulent relaxation. 

Space physics studies of relaxation have previously identified 
\citep{osman2011directional}, relaxed Alfv\'enic patches of turbulence at 1~au. Moreover, the ``patchiness'' of dynamically aligned states supports the idea of cellularized structure of magnetohydrodynamics (MHD) turbulence \citep{matthaeus2015intermittency}, the appearance of intermittency, as well as the incompatibility with the idea of turbulence arising from a superposition of Gaussian fields. The above-mentioned works investigating turbulence in the solar wind were necessarily limited by the lack of in-situ multi-spacecraft missions. Therefore, their analyses could investigate only the field alignments that do not involve the computation of derivatives e.g. Alfv\'enic states which require an (anti)alignment between the velocity and magnetic fields. Preliminary work to quantify  these correlations has been done using Cluster solar wind data \cite{servidio2014relaxation}.

The present paper closes 
a gap left in the past, namely the lack of experimental evidence assessing the potential alignments predicted by the MHD theory. This 
is made possible by employing data from the Magnetospheric Multiscale (MMS) Mission \cite{burch2016magnetospheric}. In addition, turbulence in the Earth's magnetosheath can be imagined as ``young'' \cite{matthaeus1998dynamical} since it is freshly modified by the solar wind passing through the bow shock. Identifying relaxed states in the magnetosheath strongly suggests that such states emerge quite rapidly. 


\textit{Relaxation processes in MHD.--}The relaxation processes under consideration can be studied in the context of simple MHD model equations that consist of a mix of linear and nonlinear terms. It is straightforward to show that states of minimum energy coincide with a minimization of the magnitude of the nonlinear terms. Ideal incompressible MHD equations read $\frac{\de \vv}{\de t} + \left( \vv \cdot \grad \right) \vv = - \grad p + \jj \times \bb$, $\frac{\de \bb}{\de t} = \curl ( \vv \times \bb )$, 
together with the constraints $\divg \vv = \divg \bb = 0$. In these equations, $\aa$ is the vector potential such that $\bb = \curl \aa$, $\jj = \curl \bb$ is the current density, $p$ is the kinetic pressure and $\vv$ is the velocity field.

The three quadratic rugged invariants of 3D-MHD \citep{frisch1975possibility, matthaeus1982evaluation} are the total energy $E = \frac{1}{2} \int_V \left( |\vv|^2 + |\bb|^2 \right) d^3x$, 
and the two pseudo-scalars: the cross helicity $H_c = \frac{1}{2} \int_V \vv \cdot \bb \; d^3x$, 
and the magnetic helicity $H_m = \frac{1}{2} \int_V \aa \cdot \bb \; d^3x$, 
 with the integrals being evaluated over an appropriate volume $V$.

The states that minimize energy keeping $H_c$ and $H_m$ constant can be recovered solving the variational problem $\delta \int \left(  |\vv|^2 + |\bb|^2 - 2\alpha \, \vv \cdot \bb + 2\phi \, \aa \cdot \bb  \right) d^3x = 0$
where $2\alpha$ and $2\phi$ are Lagrange multipliers. The obtained constrained fields are such that $  \vv = \alpha \bb = \frac{\alpha(1-\alpha)^2}{\phi} \jj =  \frac{(1-\alpha)^2}{\phi} \bomega$
where $\bomega = \curl \vv$ is the vorticity \cite{stribling1991relaxation, servidio2008depression, servidio2014relaxation}.

One immediately notices that the above relationships 
describe fields that tend to suppress the non-linear terms in the MHD equations. They are also associated with particular equilibrium states, such as the Taylor force-free state $\jj \propto \bb$ \citep{woltjer1958theorem,taylor1974relaxation}.
However, the reduction of nonlinearities and incomplete relaxation to the target states is of primary interest here. Moreover, suppression of nonlinearity is even more widespread; if one rewrites the term $(\vv \cdot \grad) \vv = \grad\left( \frac{v^2}{2}\right) - \vv \times \bomega$, it will be shown below that the terms on r.h.s. also evolve toward (anti)alignment. In addition, the Lorentz force term $(\jj \times \bb)$ and the Lamb vector $(\vv \times \bomega)$, each contributing to accelerations, tend to anti-align, thus 
further suppressing 
any residual nonlinearity.

A convenient measure for identifying (local) partially relaxed states is the inner product $(,)$. Given two vectors $\ff$ and $\gg$, we evaluate the distribution of the cosines of the comprised angles, namely: $\cos(\theta) = \frac{ (\ff , \gg ) }{||\ff|| \, ||\gg||}$
where $||\cdot||$ represents a vector magnitude. When $\ff$ and $\gg$ are (anti)aligned 3D vectors, the distribution of the cosine of the angles between the two tends to peak at values $\cos(\theta) = \pm 1$. On the other hand, if they have random orientations with respect to one another, the distribution will be flat at $ \cos(\theta) \sim 0.5$.

\begin{figure}[ht]
    \centering
    \includegraphics[width=0.7\columnwidth]{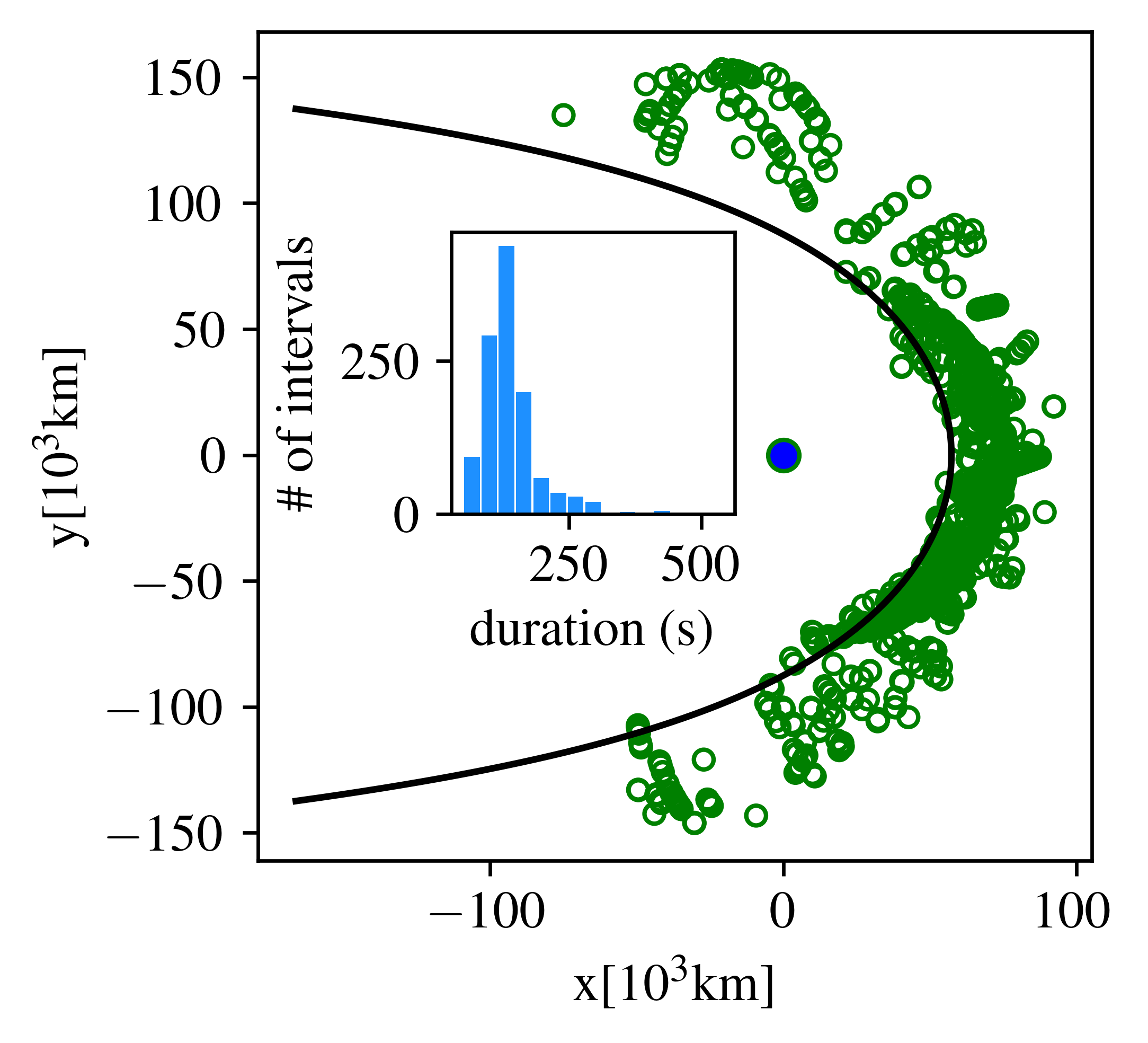}
    \caption{Nominal position of the Earth's magnetopause in GSE coordinates according to \protect\cite{shue1997new} (black curve), and locations of the MMS intervals analyzed (green circles). Earth is indicated as a blue circle. The inset shows the histogram of the durations of such intervals.}
    \label{fig:histo_duration}
\end{figure}

\textit{Data selection and analysis.--}We use a total of 1180 MMS intervals in the turbulent magnetosheath from 2015 Sep 08 through 2018 Jan 01. These range in duration from 50 to 540 seconds
as reported in Fig.~\ref{fig:histo_duration}. Velocity and magnetic field measurements are obtained from the Fast Plasma Investigation \citep[FPI,][]{pollock2016fast} and the Fluxgate magnetometers \citep[FGM,][]{russell2016magnetospheric}, respectively. Magnetic field burst measurements (128~Hz) have been resampled to match the resolution of ion bulk velocity (150~ms). Spacecraft spintone is removed from the ion data.

As a check on the FPI  data quality, we found 
density values greater than $50 \mbox{cm}^{-3}$, 
which might be less reliable. 
172 intervals out of 1180 $(\lesssim 15\%)$ have an average number density $> 50 \mbox{cm}^{-3}$. The subsequent 
analysis will not be greatly affected by these intervals.

The vorticity and current density fields are computed using the reciprocal vector technique for tetrahedra \citep{chanteur1998spatial}, resulting in a single vector time series for each. We then interpolate the bulk flow speed and the magnetic field from the four MMSs to the barycenter of the tetrahedron. Before computing the alignment cosines, it is necessary to low-pass filter all the fields to focus on inertial-range scales \cite{chhiber2018higher}. This is done by averaging over a 1-second running window, without changing the resolution. At last, the mean values are subtracted from the bulk flow and the magnetic field, since alignments are calculated between fluctuations.

\textit{Results.--}For each interval, the cosine of the angles between all the appropriate pairs of fields has been measured. As one would expect, not all intervals show indications of alignment between pairs of fields; some intervals show a good degree of relaxation, while others do not. Statistical significance is obtained by compiling distributions of the cosines of the angles of each type from all 1180 intervals used in the survey. 


\begin{figure}[ht]
    \centering
    \includegraphics[width=0.7\columnwidth]{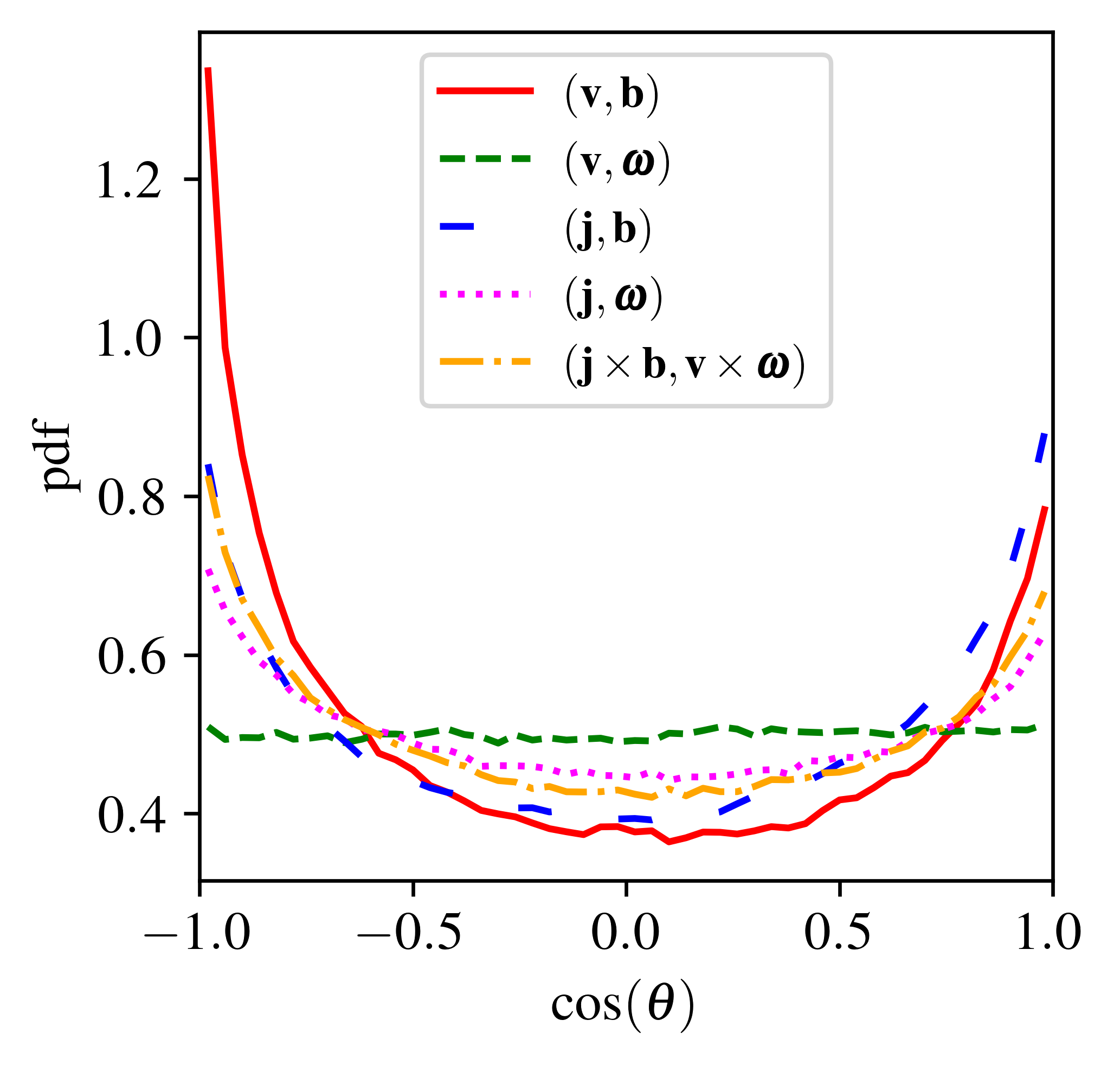}
    \caption{Global distributions of alignments evaluated over all selected intervals.}
    \label{fig:global}
\end{figure}

Figure~\ref{fig:global} shows the distribution of the alignment cosines for several relevant pairs of fields. Each  pair, when aligned (or, anti-aligned) suppresses a nonlinear term in the dynamical equations. It is apparent that there are varying degrees of alignment, and therefore varying degrees of suppression. We now briefly comment on these. 

The {\it Alfv\'enic state}, i.e., pointwise $(\vv, \bb)$ alignment, is one of the most fundamental and frequently discussed states   being associated both with linear MHD Theory and nonlinear theory. When MHD equations are written in the Els\"asser variables (defined in terms of velocity and magnetic field fluctuations $\zz^\pm = \vv \pm \bb$), it is evident that incompressible nonlinearity arises from the interaction between the $\zz^+$ and $\zz^-$ fields.  When cross helicity is high the evolution of turbulence is hampered; this requires both $(\vv, \bb)$ alignment, and energy equipartition between $(\vv$ and $\bb)$. We also investigated the behavior of $(\vv, \bb)$ alignment after sector-rectifying the magnetic field \citep{osman2011directional}, but little change was seen to occur. Although the sign of cross helicity is crucial in determining whether the Alfv\'enic fluctuations propagate outward (anti-sunward) or inward, it has no effect on the alignment. The study of the propagation of Alfv\'enic fluctuations in the magnetosheath is a problem that needs to be addressed more thoroughly \citep{yordanova2011reduced}, but such a discussion is beyond the scope of this paper.

The velocity and vorticity fields do not display global Beltrami alignment, indicated in Fig. \ref{fig:global} by $(\vv, \bomega)$. The same result was found in 3D-MHD simulations \citep{servidio2008depression} and attributed to the fact that kinetic helicity, $\int \vv \cdot \bomega \, d^3r$, is not an invariant and is therefore unconstrained. Beltrami-like alignments have also 
been previously discussed in solar wind observations \cite{servidio2014relaxation} where accuracy is limited by large spacecraft separations. 

The status of Beltrami $(\vv, \bomega)$ correlations differs in hydrodynamics (HD). A pioneering work, \cite{PelzEA85} studied velocity-vorticity patterns in turbulent flows and found that $\vv$ and $\bomega$ align in the center of a channel where the flow is relatively laminar. The alignment disappears in the turbulent transition region between the walls and the central region. When only fluctuations, and not the mean flow, were considered, alignment was lost throughout the whole channel. This parallels the present result if one imagines flux ropes to be counterparts of HD channels. A second interesting interpretation follows from the correlation between the vorticity and the rate of strain tensor and its eigenvectors. Ref. \cite{PelzEA85} found that where the dissipation (measured as $S_{ij}^2 = 1/2 (\partial u_i / \partial x_j + \partial u_j / \partial x_i )^2$) is large, the alignment is lost.

Enhancement of current-magnetic field alignment $(\jj, \bb)$ implies an approach to a force-free condition, equivalent to the Beltrami property for the magnetic field ($\nabla \times \bb \propto \bb$). Such states lend naturally to a description in terms of flux tubes, which may be locally nearly force free, but also interacting, perhaps mainly at their boundaries \cite{pecora2021identification}. When not force-free, such flux tubes may be nearly force balanced which represents an additional type of cellularization of turbulence. This type of suppression of nonlinearity, involving pressure balances, will be examined in a separate study. Here we do not involve pressure at this stage as it introduces complications such as considerate separate electron and proton pressures (separate plasma beta) \cite{Burlaga88}. Force-balanced states are often studied as Grad-Shafranov equilibria\cite{grad1958hydromagnetic}.

The $(\jj,\bomega)$ alignment could heuristically be related to the dynamics of the Alfv\'enic state, but it can also be linked to other basic properties of turbulence. In the equation that involve the generalized vorticity $\bOmega^\pm = \jj \pm \bomega$ \cite{pouquet1996turbulence}, the nonlinear term is $\zz^\pm \cdot \grad \bOmega^\mp$ that is suppressed when $\jj \sim \bomega$.

The Lorentz force and the Lamb vector show a slight anticorrelation as expected from the equations since these two terms should be of opposite signs in order to reduce any residual nonlinearity.

\begin{figure}[ht]
    \centering
    \includegraphics[width=0.7\columnwidth]{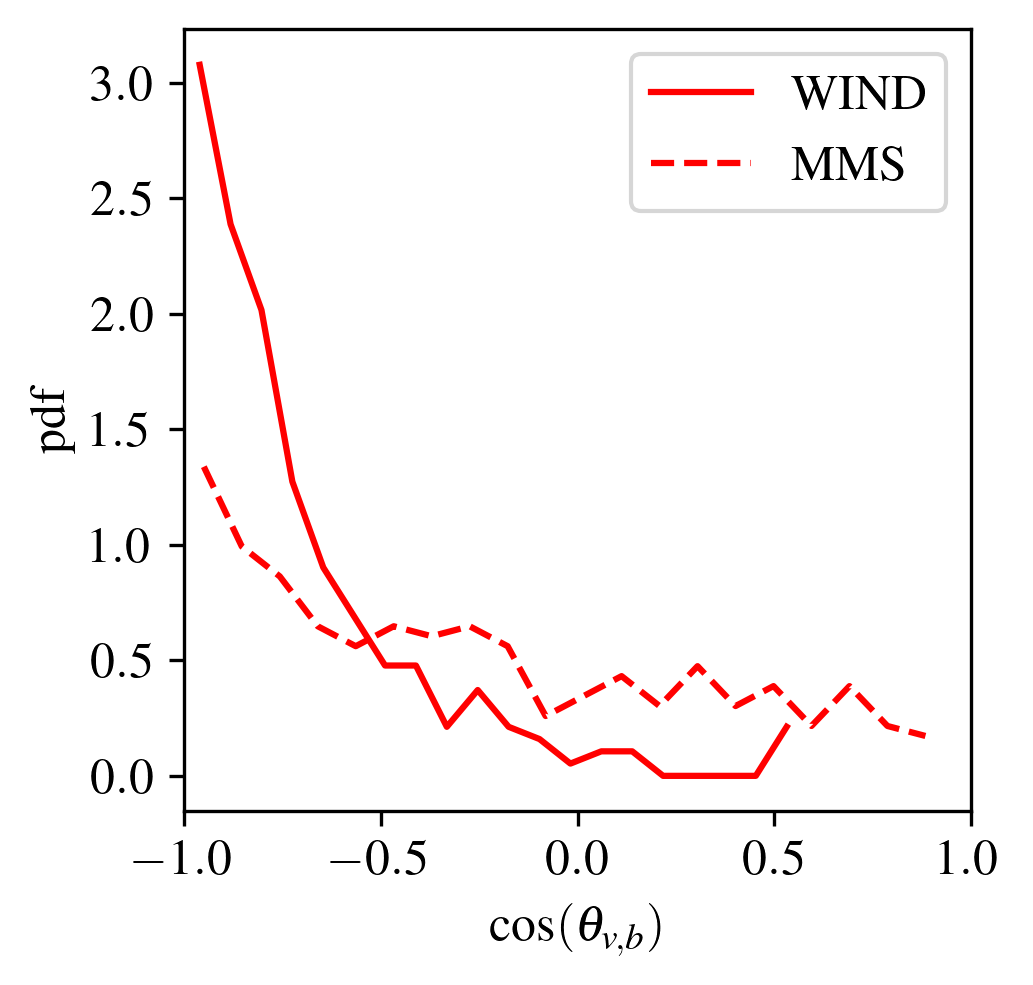}
    \caption{Alfv\'enic alignment within the same plasma parcel measured by WIND and MMS, upstream and downstream of the Earth's bow shock respectively. The distributions 
    indicate that shock crossing reduces the degree of alignment.}
    \label{fig:windmms}
\end{figure}

\textit{Transmission of relaxed patches across the bow shock.--}How turbulence is processed crossing a shock is a subject of ongoing interest \cite{ZankEA21-shocks}. In particular, a recent work \citep{trotta2022transmission} examined the possibility that flux ropes can ``survive'' the crossing of the Earth's bow shock, with different degrees of deformations. We now demonstrate how relaxed Alfv\'enic patches of solar wind plasma are modified across the bow shock. The analysis focuses on the stream of plasma that was found to be sampled by both WIND in the pristine solar wind, upstream, and, 55 minutes later, by MMS downstream of the bow shock. The interval  spans 2017 Dec 31 21:20 through 2018 Jan 01 01:20 UTC. Since Beltrami states cannot be measured by the single WIND spacecraft, we focus only on the Alfv\'enic state. We use measurements for velocity and magnetic fields from the Solar Wind Experiment \citep[SWE,][]{ogilvie1995swe} and Magnetic Field Investigation \citep[MFI,][]{lepping1995wind} suites on WIND. For consistency between the spacecraft, which have different time resolutions,  we downsample the magnetic and velocity fields to a 1-minute cadence and then evaluate the angle between the two fields for WIND and MMS separately. The two distributions of the Alfv\'enic alignment are shown in Fig.~\ref{fig:windmms}.

Clearly, the presence of alignments in the pristine solar wind does not guarantee that these will persist downstream. The evidence suggests that a small fraction of alignment is maintained, with an overall significant reduction. There is some transmission across the bow shock, but the presence of relaxed patches in the magnetosheath may also be due to the relatively rapid turbulence evolution in the magnetosheath.

\begin{figure}[ht]
    \centering
    \includegraphics[width=0.8\columnwidth]{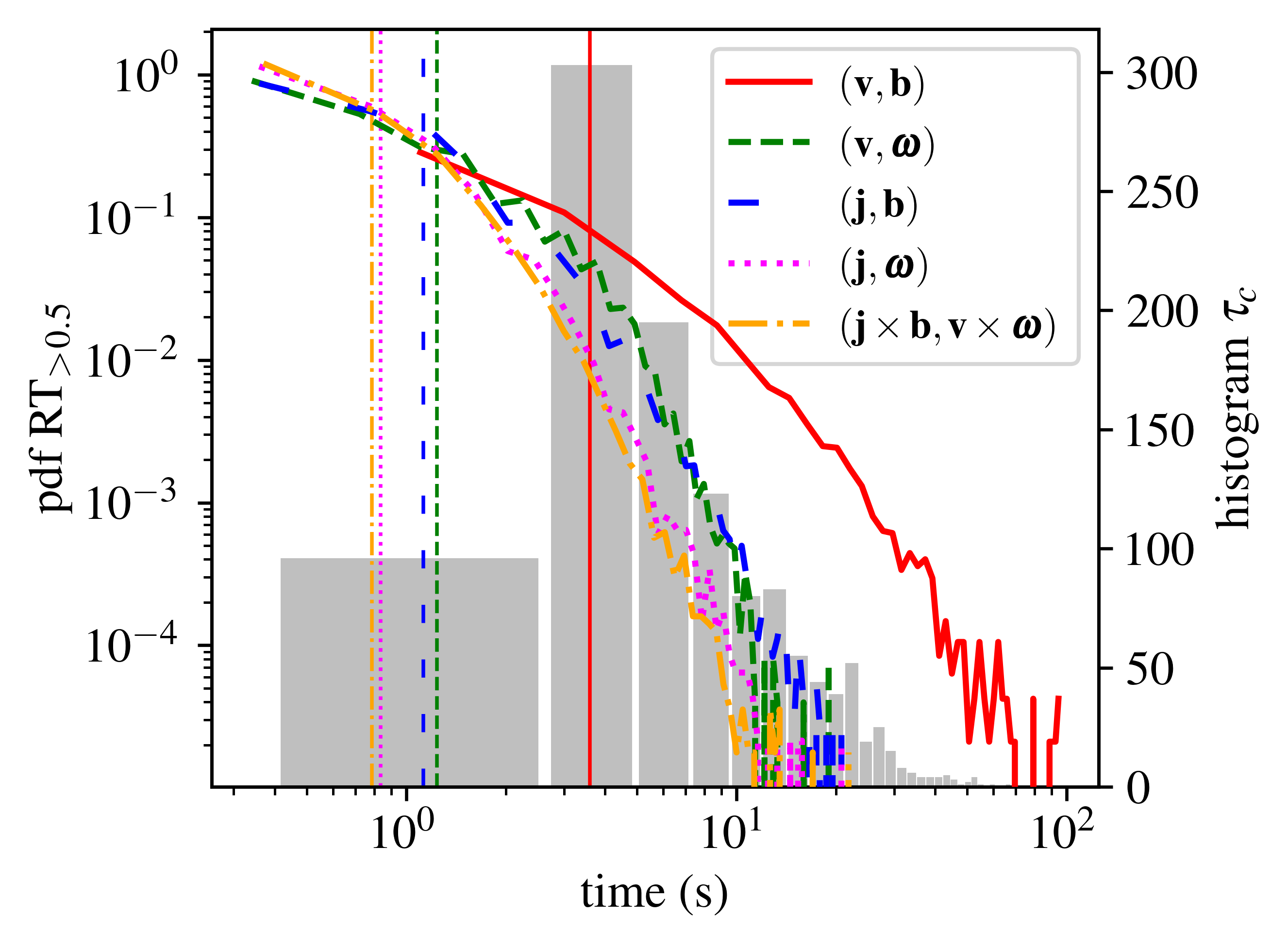}
    \caption{Probability density functions of residency times in patches where the pairs of vectors have comprised angles with a cosine larger than 0.5 or smaller than $-0.5$ (left axis). Histogram of the correlation times calculated for all the considered intervals, gray bars (right axis). Vertical lines indicate first moments of respective pdfs.}
    \label{fig:RT}
\end{figure}

\textit{Residency times in relaxed patches.--}Let us now investigate the link between the relaxed regions and the cellularization of turbulence. Starting from the cosine time series obtained as described in the previous sections, we measure for each interval how much time the system dwells above a certain alignment threshold (in absolute magnitude). The distributions of such residency times (RT) defined as cosine larger than $0.5$ or smaller than $-0.5$ are shown in Fig.~\ref{fig:RT} and indicated on the left axis as pdf RT$_{>0.5}$. The first feature one notices is that the Alfv\'en aligned states persist for longer times than the others, evidenced by a more extended distribution. The distributions show a more or less extended power law with a sharp cutoff at the upper end: This behavior is characteristic of self-similar phenomena, supporting the picture of cellularized turbulence comprising distinct patches of different sizes.

A standard estimate of the typical size of energy-containing eddies is given by the correlation scale. 
(Solar wind and magnetosheath measurements are often quoted in the time domain, the two being approximately proportional when invoking 
the Taylor frozen-in flow hypothesis \cite{jokipii1973turbulence}.) Here the eddy size and correlation scale 
are estimated using the magnetic field autocorrelation function $R(\tau)= \langle \bb(t) \cdot \bb(t+\tau) \rangle_t $, where $\bb$ are the magnetic field fluctuations and $\tau$ is the time lag. The average $\langle \dots \rangle_t$ is performed over the associated magnetosheath interval. The correlation time is determined as the first time lag at which the normalized autocorrelation function $\tilde{R}(\tau)=R(\tau)/R(0)$ falls to $1/e$ \citep[e-folding,][]{matthaeus1999correlation}. The histogram of the correlation times for each of our 1180 intervals is reported in Fig.~\ref{fig:RT}.

Finally, for each residency time distribution, we calculate the first moment (frequency of occurrence) distributions (pdfs) ${\cal{P}}(*)$ as $ T =  \int \eta \, {\cal{P}}(\eta) \, d\eta$. The probability density functions are already normalized as $ \int {\cal{P}}(\eta) \, d\eta = 1 $. In Fig.~\ref{fig:RT} the first moments T for each pdf are indicated with vertical lines having the same line style as the respective probability density.

The immediate feature one notices is the correspondence between the average residency time of the Alfv\'en state and the most probable correlation time. This correspondence is explained naturally by the above-mentioned cellularization of turbulence: The fabric of the turbulent solar wind is formed by patches (eddies) that tend toward a local equilibrium; their size can be estimated with the correlation time, and we have shown that it corresponds to the average time the magnetosheath plasma spends in an Alfv\'enic relaxed quasi-equilibrium.

The other classes of relaxed states persist, on average, for periods of time smaller than the Alfv\'enic states. There are several possible reasons for this: 
(i) non-Alfv'enic alignments involve derivatives of fields, therefore, acting on 
smaller, faster scales; 
(ii) the Alfv\'en state is more directly correlated with larger scale MHD phenomena, such as flux tubes. Additionally, the Beltrami and force-free states have similar time scales, suggesting an analogy between HD and MHD findings: From \cite{PelzEA85}, the Beltrami state is found within the central channel flow region; likewise, while in plasma, the force-free region in a flux rope may be near the magnetic axis, far from the tube's borders, where discontinuities and bursts of activity, e.g., reconnection, are often found \cite{servidio2010statistics, pecora2019single, pecora2021identification}. 

\textit{Discussion and Conclusions.--} 
The rapid appearance of characteristic correlations due to relaxation processes have been addressed previously 
from theoretical \citep{matthaeus1980selective}, numerical \citep{servidio2008depression}, and observational \citep{osman2011directional} points of view.  
We presented here, for the first time in a natural plasma, a full study 
of 
several different contributions to 
relaxation and
suppression of nonlinearity using MMS measurements in 
the magnetosheath.

Some of the special states discussed in this paper (Alfv\'enic, force-free) are well-known and studied in several different aspects in the solar wind and plasma community; though others are not. In particular, the Beltrami state, well studied in hydrodynamics, is poorly addressed in MHD and solar wind studies. The possibility that the Beltrami state is attained where the dissipation is small and destroyed where it is large may set the direction for new studies correlating dissipation and relaxation in plasma. A fundamental measure of dissipation is the pressure work \cite{yang2017energy}, which can be measured with MMS \cite{bandyopadhyay2020statistics}. Ultimately we expect that intermittency of the dissipation may be anticorrelated with relaxation, as might be inferred from flux tube structure seen in both 
simulations \cite{YangEA17-PRE} and observations \cite{BandyopadhyayEA20-PiD,pecora2021identification}.
A clear direction for future research is to examine the spatial relationship between relaxation 
signatures and dissipation in greater detail.

Regarding the fast occurrence of relaxation, previous numerical works \citep{servidio2008depression} assessed the emergence of alignments after a few Alfv\'en times. However, the related observational studies performed in the solar wind \citep{osman2011directional, servidio2014relaxation} could not infer any temporal evolution, as 1~au turbulence potentially had at least several nonlinear/Alfv\'en times to evolve \cite{matthaeus1998dynamical}. Conversely, the magnetosheath plasma consists of strongly perturbed solar wind plasma (by passage through the bow shock) while it 
also evolves on much more rapid timescales \cite{StawarzEA22}. 
Therefore, the presence of relaxed patches in the magnetosheath suggests 
that relaxation is a fast-occurring phenomenon, but it is also possible 
that relaxed patches are degraded but still survive passage through the shock. 
It is worth mentioning the recent results from \citep{trotta2022transmission} that investigated the transmission of flux ropes across the Earth's bow shock. Their finding is that some structures can survive the crossing of the bow shock, hinting that relaxed patches downstream may come directly from the solar wind (possibly with a reduced degree of alignment.)

In hydrodynamics \cite{kerr1985higher, blackburn1996topology, pouquet1996turbulence}
relaxation often involves alignments of eigenvectors of the rate-of-strain tensor with other quantities, such as vorticity. 
Such alignments provide additional topological information. 
Such study, adapted to the  turbulent fields in plasmas \cite{consolini2015statistics}
may provide valuable physical insights in the 
study of 
magnetosheath turbulence. 

By examining relaxation in the magnetosheath, this study closes a gap in the current literature. 
We also expect that it will motivate a number of additional studies as suggested above.
A major extension to the
solar wind awaits upcoming multispacecraft missions such as Helioswarm \cite{SpenceEA19}.

\begin{acknowledgments}
This research is supported in part by the MMS Theory and Modeling program grant 80NSSC19K0284,
 and the NSF/DOE program under grant AGS-2108834 at the University of Delaware.
\end{acknowledgments}


\begin{thebibliography}{63}%
\makeatletter
\providecommand \@ifxundefined [1]{%
 \@ifx{#1\undefined}
}%
\providecommand \@ifnum [1]{%
 \ifnum #1\expandafter \@firstoftwo
 \else \expandafter \@secondoftwo
 \fi
}%
\providecommand \@ifx [1]{%
 \ifx #1\expandafter \@firstoftwo
 \else \expandafter \@secondoftwo
 \fi
}%
\providecommand \natexlab [1]{#1}%
\providecommand \enquote  [1]{``#1''}%
\providecommand \bibnamefont  [1]{#1}%
\providecommand \bibfnamefont [1]{#1}%
\providecommand \citenamefont [1]{#1}%
\providecommand \href@noop [0]{\@secondoftwo}%
\providecommand \href [0]{\begingroup \@sanitize@url \@href}%
\providecommand \@href[1]{\@@startlink{#1}\@@href}%
\providecommand \@@href[1]{\endgroup#1\@@endlink}%
\providecommand \@sanitize@url [0]{\catcode `\\12\catcode `\$12\catcode
  `\&12\catcode `\#12\catcode `\^12\catcode `\_12\catcode `\%12\relax}%
\providecommand \@@startlink[1]{}%
\providecommand \@@endlink[0]{}%
\providecommand \url  [0]{\begingroup\@sanitize@url \@url }%
\providecommand \@url [1]{\endgroup\@href {#1}{\urlprefix }}%
\providecommand \urlprefix  [0]{URL }%
\providecommand \Eprint [0]{\href }%
\providecommand \doibase [0]{https://doi.org/}%
\providecommand \selectlanguage [0]{\@gobble}%
\providecommand \bibinfo  [0]{\@secondoftwo}%
\providecommand \bibfield  [0]{\@secondoftwo}%
\providecommand \translation [1]{[#1]}%
\providecommand \BibitemOpen [0]{}%
\providecommand \bibitemStop [0]{}%
\providecommand \bibitemNoStop [0]{.\EOS\space}%
\providecommand \EOS [0]{\spacefactor3000\relax}%
\providecommand \BibitemShut  [1]{\csname bibitem#1\endcsname}%
\let\auto@bib@innerbib\@empty
\bibitem [{\citenamefont {Matthaeus}\ \emph {et~al.}(2012)\citenamefont
  {Matthaeus}, \citenamefont {Montgomery}, \citenamefont {Wan},\ and\
  \citenamefont {Servidio}}]{matthaeus2012review}%
  \BibitemOpen
  \bibfield  {author} {\bibinfo {author} {\bibfnamefont {W.~H.}\ \bibnamefont
  {Matthaeus}}, \bibinfo {author} {\bibfnamefont {D.~C.}\ \bibnamefont
  {Montgomery}}, \bibinfo {author} {\bibfnamefont {M.}~\bibnamefont {Wan}},\
  and\ \bibinfo {author} {\bibfnamefont {S.}~\bibnamefont {Servidio}},\
  }\bibfield  {title} {\bibinfo {title} {A review of relaxation and structure
  in some turbulent plasmas: magnetohydrodynamics and related models},\ }\href
  {https://doi.org/10.1080/14685248.2012.704378} {\bibfield  {journal}
  {\bibinfo  {journal} {Journal of Turbulence}\ }\textbf {\bibinfo {volume}
  {13}},\ \bibinfo {pages} {N37} (\bibinfo {year} {2012})}\BibitemShut
  {NoStop}%
\bibitem [{\citenamefont {Frisch}\ \emph {et~al.}(1975)\citenamefont {Frisch},
  \citenamefont {Pouquet}, \citenamefont {LÉOrat},\ and\ \citenamefont
  {Mazure}}]{frisch1975possibility}%
  \BibitemOpen
  \bibfield  {author} {\bibinfo {author} {\bibfnamefont {U.}~\bibnamefont
  {Frisch}}, \bibinfo {author} {\bibfnamefont {A.}~\bibnamefont {Pouquet}},
  \bibinfo {author} {\bibfnamefont {J.}~\bibnamefont {LÉOrat}},\ and\ \bibinfo
  {author} {\bibfnamefont {A.}~\bibnamefont {Mazure}},\ }\bibfield  {title}
  {\bibinfo {title} {Possibility of an inverse cascade of magnetic helicity in
  magnetohydrodynamic turbulence},\ }\href
  {https://doi.org/10.1017/S002211207500122X} {\bibfield  {journal} {\bibinfo
  {journal} {Journal of Fluid Mechanics}\ }\textbf {\bibinfo {volume} {68}},\
  \bibinfo {pages} {769–778} (\bibinfo {year} {1975})}\BibitemShut {NoStop}%
\bibitem [{\citenamefont {Montgomery}\ \emph {et~al.}(1978)\citenamefont
  {Montgomery}, \citenamefont {Turner},\ and\ \citenamefont
  {Vahala}}]{montgomery1978three}%
  \BibitemOpen
  \bibfield  {author} {\bibinfo {author} {\bibfnamefont {D.}~\bibnamefont
  {Montgomery}}, \bibinfo {author} {\bibfnamefont {L.}~\bibnamefont {Turner}},\
  and\ \bibinfo {author} {\bibfnamefont {G.}~\bibnamefont {Vahala}},\
  }\bibfield  {title} {\bibinfo {title} {Three‐dimensional
  magnetohydrodynamic turbulence in cylindrical geometry},\ }\href
  {https://doi.org/10.1063/1.862295} {\bibfield  {journal} {\bibinfo  {journal}
  {The Physics of Fluids}\ }\textbf {\bibinfo {volume} {21}},\ \bibinfo {pages}
  {757} (\bibinfo {year} {1978})}\BibitemShut {NoStop}%
\bibitem [{\citenamefont {Stribling}\ and\ \citenamefont
  {Matthaeus}(1990)}]{stribling1991statistical}%
  \BibitemOpen
  \bibfield  {author} {\bibinfo {author} {\bibfnamefont {T.}~\bibnamefont
  {Stribling}}\ and\ \bibinfo {author} {\bibfnamefont {W.~H.}\ \bibnamefont
  {Matthaeus}},\ }\bibfield  {title} {\bibinfo {title} {Statistical properties
  of ideal three‐dimensional magnetohydrodynamics},\ }\href
  {https://doi.org/10.1063/1.859419} {\bibfield  {journal} {\bibinfo  {journal}
  {Physics of Fluids B: Plasma Physics}\ }\textbf {\bibinfo {volume} {2}},\
  \bibinfo {pages} {1979} (\bibinfo {year} {1990})}\BibitemShut {NoStop}%
\bibitem [{\citenamefont {Matthaeus}\ and\ \citenamefont
  {Montgomery}(1980{\natexlab{a}})}]{matthaeus1980selective}%
  \BibitemOpen
  \bibfield  {author} {\bibinfo {author} {\bibfnamefont {W.~H.}\ \bibnamefont
  {Matthaeus}}\ and\ \bibinfo {author} {\bibfnamefont {D.}~\bibnamefont
  {Montgomery}},\ }\bibfield  {title} {\bibinfo {title} {Selective decay
  hypothesis at high mechanical and magnetic reynolds numbers},\ }\href
  {https://doi.org/10.1111/j.1749-6632.1980.tb29687.x} {\bibfield  {journal}
  {\bibinfo  {journal} {New York Academy of Sciences, Annals}\ }\textbf
  {\bibinfo {volume} {357}},\ \bibinfo {pages} {203} (\bibinfo {year}
  {1980}{\natexlab{a}})}\BibitemShut {NoStop}%
\bibitem [{\citenamefont {Ting}\ \emph {et~al.}(1986)\citenamefont {Ting},
  \citenamefont {Matthaeus},\ and\ \citenamefont
  {Montgomery}}]{ting1986turbulent}%
  \BibitemOpen
  \bibfield  {author} {\bibinfo {author} {\bibfnamefont {A.~C.}\ \bibnamefont
  {Ting}}, \bibinfo {author} {\bibfnamefont {W.~H.}\ \bibnamefont
  {Matthaeus}},\ and\ \bibinfo {author} {\bibfnamefont {D.}~\bibnamefont
  {Montgomery}},\ }\bibfield  {title} {\bibinfo {title} {Turbulent relaxation
  processes in magnetohydrodynamics},\ }\href
  {https://doi.org/10.1063/1.865843} {\bibfield  {journal} {\bibinfo  {journal}
  {The Physics of Fluids}\ }\textbf {\bibinfo {volume} {29}},\ \bibinfo {pages}
  {3261} (\bibinfo {year} {1986})}\BibitemShut {NoStop}%
\bibitem [{\citenamefont {Stribling}\ and\ \citenamefont
  {Matthaeus}(1991)}]{stribling1991relaxation}%
  \BibitemOpen
  \bibfield  {author} {\bibinfo {author} {\bibfnamefont {T.}~\bibnamefont
  {Stribling}}\ and\ \bibinfo {author} {\bibfnamefont {W.~H.}\ \bibnamefont
  {Matthaeus}},\ }\bibfield  {title} {\bibinfo {title} {Relaxation processes in
  a low‐order three‐dimensional magnetohydrodynamics model},\ }\href
  {https://doi.org/10.1063/1.859654} {\bibfield  {journal} {\bibinfo  {journal}
  {Physics of Fluids B: Plasma Physics}\ }\textbf {\bibinfo {volume} {3}},\
  \bibinfo {pages} {1848} (\bibinfo {year} {1991})},\ \Eprint
  {https://arxiv.org/abs/https://doi.org/10.1063/1.859654}
  {https://doi.org/10.1063/1.859654} \BibitemShut {NoStop}%
\bibitem [{\citenamefont {{Pelz}}\ \emph {et~al.}(1985)\citenamefont {{Pelz}},
  \citenamefont {{Yakhot}}, \citenamefont {{Orszag}}, \citenamefont
  {{Shtilman}},\ and\ \citenamefont {{Levich}}}]{PelzEA85}%
  \BibitemOpen
  \bibfield  {author} {\bibinfo {author} {\bibfnamefont {R.~B.}\ \bibnamefont
  {{Pelz}}}, \bibinfo {author} {\bibfnamefont {V.}~\bibnamefont {{Yakhot}}},
  \bibinfo {author} {\bibfnamefont {S.~A.}\ \bibnamefont {{Orszag}}}, \bibinfo
  {author} {\bibfnamefont {L.}~\bibnamefont {{Shtilman}}},\ and\ \bibinfo
  {author} {\bibfnamefont {E.}~\bibnamefont {{Levich}}},\ }\bibfield  {title}
  {\bibinfo {title} {{Velocity-vorticity patterns in turbulent flow}},\
  }\href@noop {} {\bibfield  {journal} {\bibinfo  {journal} {Physical Review
  Letters}\ }\textbf {\bibinfo {volume} {54}},\ \bibinfo {pages} {2505}
  (\bibinfo {year} {1985})}\BibitemShut {NoStop}%
\bibitem [{\citenamefont {Montgomery}\ \emph {et~al.}(1992)\citenamefont
  {Montgomery}, \citenamefont {Matthaeus}, \citenamefont {Stribling},
  \citenamefont {Mart\'{\i}nez},\ and\ \citenamefont {Oughton}}]{MontEA92-pof}%
  \BibitemOpen
  \bibfield  {author} {\bibinfo {author} {\bibfnamefont {D.~C.}\ \bibnamefont
  {Montgomery}}, \bibinfo {author} {\bibfnamefont {W.~H.}\ \bibnamefont
  {Matthaeus}}, \bibinfo {author} {\bibfnamefont {W.~T.}\ \bibnamefont
  {Stribling}}, \bibinfo {author} {\bibfnamefont {D.}~\bibnamefont
  {Mart\'{\i}nez}},\ and\ \bibinfo {author} {\bibfnamefont {S.}~\bibnamefont
  {Oughton}},\ }\bibfield  {title} {\bibinfo {title} {Relaxation in two
  dimensions and the ``sinh-{Poisson}'' equation},\ }\href
  {https://doi.org/10.1063/1.858525} {\bibfield  {journal} {\bibinfo  {journal}
  {Phys.\ Fluids A}\ }\textbf {\bibinfo {volume} {4}},\ \bibinfo {pages} {3}
  (\bibinfo {year} {1992})}\BibitemShut {NoStop}%
\bibitem [{\citenamefont {Servidio}\ \emph {et~al.}(2008)\citenamefont
  {Servidio}, \citenamefont {Matthaeus},\ and\ \citenamefont
  {Dmitruk}}]{servidio2008depression}%
  \BibitemOpen
  \bibfield  {author} {\bibinfo {author} {\bibfnamefont {S.}~\bibnamefont
  {Servidio}}, \bibinfo {author} {\bibfnamefont {W.~H.}\ \bibnamefont
  {Matthaeus}},\ and\ \bibinfo {author} {\bibfnamefont {P.}~\bibnamefont
  {Dmitruk}},\ }\bibfield  {title} {\bibinfo {title} {Depression of
  nonlinearity in decaying isotropic mhd turbulence},\ }\href
  {https://doi.org/10.1103/PhysRevLett.100.095005} {\bibfield  {journal}
  {\bibinfo  {journal} {Phys. Rev. Lett.}\ }\textbf {\bibinfo {volume} {100}},\
  \bibinfo {pages} {095005} (\bibinfo {year} {2008})}\BibitemShut {NoStop}%
\bibitem [{\citenamefont {Matthaeus}\ \emph {et~al.}(2008)\citenamefont
  {Matthaeus}, \citenamefont {Pouquet}, \citenamefont {Mininni}, \citenamefont
  {Dmitruk},\ and\ \citenamefont {Breech}}]{MattEA08-align}%
  \BibitemOpen
  \bibfield  {author} {\bibinfo {author} {\bibfnamefont {W.~H.}\ \bibnamefont
  {Matthaeus}}, \bibinfo {author} {\bibfnamefont {A.}~\bibnamefont {Pouquet}},
  \bibinfo {author} {\bibfnamefont {P.~D.}\ \bibnamefont {Mininni}}, \bibinfo
  {author} {\bibfnamefont {P.}~\bibnamefont {Dmitruk}},\ and\ \bibinfo {author}
  {\bibfnamefont {B.}~\bibnamefont {Breech}},\ }\bibfield  {title} {\bibinfo
  {title} {Rapid alignment of velocity and magnetic field in
  magnetohydrodynamic turbulence},\ }\href
  {https://doi.org/10.1103/PhysRevLett.100.085003} {\bibfield  {journal}
  {\bibinfo  {journal} {Phys.\ Rev.\ Lett.}\ }\textbf {\bibinfo {volume}
  {100}},\ \bibinfo {eid} {085003} (\bibinfo {year} {2008})}\BibitemShut
  {NoStop}%
\bibitem [{\citenamefont {{Servidio}}\ \emph {et~al.}(2010)\citenamefont
  {{Servidio}}, \citenamefont {{Wan}}, \citenamefont {{Matthaeus}},\ and\
  \citenamefont {{Carbone}}}]{ServidioEA10-PF}%
  \BibitemOpen
  \bibfield  {author} {\bibinfo {author} {\bibfnamefont {S.}~\bibnamefont
  {{Servidio}}}, \bibinfo {author} {\bibfnamefont {M.}~\bibnamefont {{Wan}}},
  \bibinfo {author} {\bibfnamefont {W.~H.}\ \bibnamefont {{Matthaeus}}},\ and\
  \bibinfo {author} {\bibfnamefont {V.}~\bibnamefont {{Carbone}}},\ }\bibfield
  {title} {\bibinfo {title} {{Local relaxation and maximum entropy in
  two-dimensional turbulence}},\ }\href {https://doi.org/10.1063/1.3526760}
  {\bibfield  {journal} {\bibinfo  {journal} {Physics of Fluids}\ }\textbf
  {\bibinfo {volume} {22}},\ \bibinfo {pages} {125107} (\bibinfo {year}
  {2010})}\BibitemShut {NoStop}%
\bibitem [{\citenamefont {Montgomery}\ and\ \citenamefont
  {Joyce}(1974)}]{MontJoyce74}%
  \BibitemOpen
  \bibfield  {author} {\bibinfo {author} {\bibfnamefont {D.~C.}\ \bibnamefont
  {Montgomery}}\ and\ \bibinfo {author} {\bibfnamefont {G.}~\bibnamefont
  {Joyce}},\ }\bibfield  {title} {\bibinfo {title} {Statistical mechanics of
  ``negative temperature'' states},\ }\href {https://doi.org/10.1063/1.1694856}
  {\bibfield  {journal} {\bibinfo  {journal} {Phys.\ Fluids}\ }\textbf
  {\bibinfo {volume} {17}},\ \bibinfo {pages} {1139} (\bibinfo {year}
  {1974})}\BibitemShut {NoStop}%
\bibitem [{\citenamefont {Montgomery}\ \emph {et~al.}(1993)\citenamefont
  {Montgomery}, \citenamefont {Shan},\ and\ \citenamefont
  {Matthaeus}}]{MontEA93}%
  \BibitemOpen
  \bibfield  {author} {\bibinfo {author} {\bibfnamefont {D.~C.}\ \bibnamefont
  {Montgomery}}, \bibinfo {author} {\bibfnamefont {X.}~\bibnamefont {Shan}},\
  and\ \bibinfo {author} {\bibfnamefont {W.~H.}\ \bibnamefont {Matthaeus}},\
  }\bibfield  {title} {\bibinfo {title} {Navier--{Stokes} relaxation to
  sinh-{Poisson} states at finite {Reynolds} numbers},\ }\href@noop {}
  {\bibfield  {journal} {\bibinfo  {journal} {Phys.\ Fluids A}\ }\textbf
  {\bibinfo {volume} {5}},\ \bibinfo {pages} {2207} (\bibinfo {year}
  {1993})}\BibitemShut {NoStop}%
\bibitem [{\citenamefont {Servidio}\ \emph
  {et~al.}(2010{\natexlab{a}})\citenamefont {Servidio}, \citenamefont {Wan},
  \citenamefont {Matthaeus},\ and\ \citenamefont
  {Carbone}}]{ServidioEA10-2Dns}%
  \BibitemOpen
  \bibfield  {author} {\bibinfo {author} {\bibfnamefont {S.}~\bibnamefont
  {Servidio}}, \bibinfo {author} {\bibfnamefont {M.}~\bibnamefont {Wan}},
  \bibinfo {author} {\bibfnamefont {W.~H.}\ \bibnamefont {Matthaeus}},\ and\
  \bibinfo {author} {\bibfnamefont {V.}~\bibnamefont {Carbone}},\ }\bibfield
  {title} {\bibinfo {title} {Local relaxation and maximum entropy in
  two-dimensional turbulence},\ }\href {https://doi.org/10.1063/1.3526760}
  {\bibfield  {journal} {\bibinfo  {journal} {Phys.\ Fluids}\ }\textbf
  {\bibinfo {volume} {22}},\ \bibinfo {eid} {125107} (\bibinfo {year}
  {2010}{\natexlab{a}})}\BibitemShut {NoStop}%
\bibitem [{\citenamefont {Rodgers}\ \emph {et~al.}(2009)\citenamefont
  {Rodgers}, \citenamefont {Servidio}, \citenamefont {Matthaeus}, \citenamefont
  {Montgomery}, \citenamefont {Mitchell},\ and\ \citenamefont
  {Aziz}}]{RodgersEA09}%
  \BibitemOpen
  \bibfield  {author} {\bibinfo {author} {\bibfnamefont {D.~J.}\ \bibnamefont
  {Rodgers}}, \bibinfo {author} {\bibfnamefont {S.}~\bibnamefont {Servidio}},
  \bibinfo {author} {\bibfnamefont {W.~H.}\ \bibnamefont {Matthaeus}}, \bibinfo
  {author} {\bibfnamefont {D.~C.}\ \bibnamefont {Montgomery}}, \bibinfo
  {author} {\bibfnamefont {T.~B.}\ \bibnamefont {Mitchell}},\ and\ \bibinfo
  {author} {\bibfnamefont {T.}~\bibnamefont {Aziz}},\ }\bibfield  {title}
  {\bibinfo {title} {Hydrodynamic relaxation of an electron plasma to a
  near-maximum entropy state},\ }\href
  {https://doi.org/10.1103/PhysRevLett.102.244501} {\bibfield  {journal}
  {\bibinfo  {journal} {Phys.\ Rev.\ Lett.}\ }\textbf {\bibinfo {volume}
  {102}},\ \bibinfo {eid} {244501} (\bibinfo {year} {2009})}\BibitemShut
  {NoStop}%
\bibitem [{\citenamefont {Kintner}\ and\ \citenamefont
  {Seyler}(1985)}]{KintnerKelley85}%
  \BibitemOpen
  \bibfield  {author} {\bibinfo {author} {\bibfnamefont {P.~M.}\ \bibnamefont
  {Kintner}}\ and\ \bibinfo {author} {\bibfnamefont {C.~E.}\ \bibnamefont
  {Seyler}},\ }\bibfield  {title} {\bibinfo {title} {The status of observations
  and theory of high latitude ionospheric and magnetospheric plasma
  turbulence},\ }\href@noop {} {\bibfield  {journal} {\bibinfo  {journal}
  {Space Science Reviews}\ }\textbf {\bibinfo {volume} {41}},\ \bibinfo {pages}
  {91} (\bibinfo {year} {1985})}\BibitemShut {NoStop}%
\bibitem [{\citenamefont {Bouchet}\ and\ \citenamefont
  {Venaille}(2012)}]{BouchetVenaille12}%
  \BibitemOpen
  \bibfield  {author} {\bibinfo {author} {\bibfnamefont {F.}~\bibnamefont
  {Bouchet}}\ and\ \bibinfo {author} {\bibfnamefont {A.}~\bibnamefont
  {Venaille}},\ }\bibfield  {title} {\bibinfo {title} {Statistical mechanics of
  two-dimensional and geophysical flows},\ }\href
  {https://doi.org/10.1016/j.physrep.2012.02.001} {\bibfield  {journal}
  {\bibinfo  {journal} {Phys.\ Rep.}\ }\textbf {\bibinfo {volume} {515}},\
  \bibinfo {pages} {227} (\bibinfo {year} {2012})}\BibitemShut {NoStop}%
\bibitem [{\citenamefont {Matthaeus}\ and\ \citenamefont
  {Montgomery}(1980{\natexlab{b}})}]{MattMont80}%
  \BibitemOpen
  \bibfield  {author} {\bibinfo {author} {\bibfnamefont {W.~H.}\ \bibnamefont
  {Matthaeus}}\ and\ \bibinfo {author} {\bibfnamefont {D.}~\bibnamefont
  {Montgomery}},\ }\bibfield  {title} {\bibinfo {title} {Selective decay
  hypothesis at high mechanical and magnetic {R}eynolds numbers},\ }\href
  {https://doi.org/10.1111/j.1749-6632.1980.tb29687.x} {\bibfield  {journal}
  {\bibinfo  {journal} {Annals of the New York Academy of Sciences}\ }\textbf
  {\bibinfo {volume} {357}},\ \bibinfo {pages} {203} (\bibinfo {year}
  {1980}{\natexlab{b}})}\BibitemShut {NoStop}%
\bibitem [{\citenamefont {Chandrasekhar}\ and\ \citenamefont
  {Woltjer}(1958)}]{chandrasekhar1958force}%
  \BibitemOpen
  \bibfield  {author} {\bibinfo {author} {\bibfnamefont {S.}~\bibnamefont
  {Chandrasekhar}}\ and\ \bibinfo {author} {\bibfnamefont {L.}~\bibnamefont
  {Woltjer}},\ }\bibfield  {title} {\bibinfo {title} {On force-free magnetic
  fields},\ }\href {https://doi.org/10.1073/pnas.44.4.285} {\bibfield
  {journal} {\bibinfo  {journal} {Proceedings of the National Academy of
  Sciences}\ }\textbf {\bibinfo {volume} {44}},\ \bibinfo {pages} {285}
  (\bibinfo {year} {1958})}\BibitemShut {NoStop}%
\bibitem [{\citenamefont {{Taylor}}(1974)}]{taylor1974relaxation}%
  \BibitemOpen
  \bibfield  {author} {\bibinfo {author} {\bibfnamefont {J.~B.}\ \bibnamefont
  {{Taylor}}},\ }\bibfield  {title} {\bibinfo {title} {{Relaxation of Toroidal
  Plasma and Generation of Reverse Magnetic Fields}},\ }\href
  {https://doi.org/10.1103/PhysRevLett.33.1139} {\bibfield  {journal} {\bibinfo
   {journal} {\prl}\ }\textbf {\bibinfo {volume} {33}},\ \bibinfo {pages}
  {1139} (\bibinfo {year} {1974})}\BibitemShut {NoStop}%
\bibitem [{\citenamefont {{Taylor}}(1986)}]{Taylor86}%
  \BibitemOpen
  \bibfield  {author} {\bibinfo {author} {\bibfnamefont {J.~B.}\ \bibnamefont
  {{Taylor}}},\ }\bibfield  {title} {\bibinfo {title} {{Relaxation and magnetic
  reconnection in plasmas}},\ }\href
  {https://doi.org/10.1103/RevModPhys.58.741} {\bibfield  {journal} {\bibinfo
  {journal} {Reviews of Modern Physics}\ }\textbf {\bibinfo {volume} {58}},\
  \bibinfo {pages} {741} (\bibinfo {year} {1986})}\BibitemShut {NoStop}%
\bibitem [{\citenamefont
  {Woltjer}(1958{\natexlab{a}})}]{woltjer1958hydromagnetic}%
  \BibitemOpen
  \bibfield  {author} {\bibinfo {author} {\bibfnamefont {L.}~\bibnamefont
  {Woltjer}},\ }\bibfield  {title} {\bibinfo {title} {On hydromagnetic
  equilibrium},\ }\href {https://doi.org/10.1073/pnas.44.9.833} {\bibfield
  {journal} {\bibinfo  {journal} {Proceedings of the National Academy of
  Sciences}\ }\textbf {\bibinfo {volume} {44}},\ \bibinfo {pages} {833}
  (\bibinfo {year} {1958}{\natexlab{a}})}\BibitemShut {NoStop}%
\bibitem [{\citenamefont {Pouquet}\ \emph {et~al.}(1986)\citenamefont
  {Pouquet}, \citenamefont {Meneguzzi},\ and\ \citenamefont
  {Frisch}}]{PouquetEA86}%
  \BibitemOpen
  \bibfield  {author} {\bibinfo {author} {\bibfnamefont {A.}~\bibnamefont
  {Pouquet}}, \bibinfo {author} {\bibfnamefont {M.}~\bibnamefont {Meneguzzi}},\
  and\ \bibinfo {author} {\bibfnamefont {U.}~\bibnamefont {Frisch}},\
  }\bibfield  {title} {\bibinfo {title} {Growth of correlations in
  magnetohydrodynamic turbulence},\ }\href
  {https://doi.org/10.1103/PhysRevA.33.4266} {\bibfield  {journal} {\bibinfo
  {journal} {Phys.\ Rev.\ A}\ }\textbf {\bibinfo {volume} {33}},\ \bibinfo
  {pages} {4266} (\bibinfo {year} {1986})}\BibitemShut {NoStop}%
\bibitem [{\citenamefont {{Burlaga}}\ \emph {et~al.}(1981)\citenamefont
  {{Burlaga}}, \citenamefont {{Sittler}}, \citenamefont {{Mariani}},\ and\
  \citenamefont {{Schwenn}}}]{burlaga1981magnetic}%
  \BibitemOpen
  \bibfield  {author} {\bibinfo {author} {\bibfnamefont {L.}~\bibnamefont
  {{Burlaga}}}, \bibinfo {author} {\bibfnamefont {E.}~\bibnamefont
  {{Sittler}}}, \bibinfo {author} {\bibfnamefont {F.}~\bibnamefont
  {{Mariani}}},\ and\ \bibinfo {author} {\bibfnamefont {R.}~\bibnamefont
  {{Schwenn}}},\ }\bibfield  {title} {\bibinfo {title} {{Magnetic loop behind
  an interplanetary shock: Voyager, Helios, and IMP 8 observations}},\ }\href
  {https://doi.org/10.1029/JA086iA08p06673} {\bibfield  {journal} {\bibinfo
  {journal} {\jgr}\ }\textbf {\bibinfo {volume} {86}},\ \bibinfo {pages} {6673}
  (\bibinfo {year} {1981})}\BibitemShut {NoStop}%
\bibitem [{\citenamefont {Heyvaerts}\ and\ \citenamefont
  {Priest}(1984)}]{HeyvaertsPriest84}%
  \BibitemOpen
  \bibfield  {author} {\bibinfo {author} {\bibfnamefont {J.}~\bibnamefont
  {Heyvaerts}}\ and\ \bibinfo {author} {\bibfnamefont {E.~R.}\ \bibnamefont
  {Priest}},\ }\bibfield  {title} {\bibinfo {title} {Coronal heating by
  reconnection in {D}{C} current systems. {A} theory based on {T}aylor's
  hypothesis},\ }\href@noop {} {\bibfield  {journal} {\bibinfo  {journal}
  {Astron.\ Astrophys.}\ }\textbf {\bibinfo {volume} {137}},\ \bibinfo {pages}
  {63} (\bibinfo {year} {1984})}\BibitemShut {NoStop}%
\bibitem [{\citenamefont {Grad}\ and\ \citenamefont
  {Rubin}(1958)}]{grad1958hydromagnetic}%
  \BibitemOpen
  \bibfield  {author} {\bibinfo {author} {\bibfnamefont {H.}~\bibnamefont
  {Grad}}\ and\ \bibinfo {author} {\bibfnamefont {H.}~\bibnamefont {Rubin}},\
  }\bibfield  {title} {\bibinfo {title} {Hydromagnetic equilibria and
  force-free fields},\ }\href@noop {} {\bibfield  {journal} {\bibinfo
  {journal} {Journal of Nuclear Energy (1954)}\ }\textbf {\bibinfo {volume}
  {7}},\ \bibinfo {pages} {284} (\bibinfo {year} {1958})}\BibitemShut {NoStop}%
\bibitem [{\citenamefont {{Sonnerup}}\ \emph {et~al.}(2016)\citenamefont
  {{Sonnerup}}, \citenamefont {{Hasegawa}}, \citenamefont {{Denton}},\ and\
  \citenamefont {{Nakamura}}}]{sonnerup2016reconstruction}%
  \BibitemOpen
  \bibfield  {author} {\bibinfo {author} {\bibfnamefont {B.~U.~{\"O}.}\
  \bibnamefont {{Sonnerup}}}, \bibinfo {author} {\bibfnamefont
  {H.}~\bibnamefont {{Hasegawa}}}, \bibinfo {author} {\bibfnamefont {R.~E.}\
  \bibnamefont {{Denton}}},\ and\ \bibinfo {author} {\bibfnamefont {T.~K.~M.}\
  \bibnamefont {{Nakamura}}},\ }\bibfield  {title} {\bibinfo {title}
  {{Reconstruction of the electron diffusion region}},\ }\href
  {https://doi.org/10.1002/2016JA022430} {\bibfield  {journal} {\bibinfo
  {journal} {Journal of Geophysical Research (Space Physics)}\ }\textbf
  {\bibinfo {volume} {121}},\ \bibinfo {pages} {4279} (\bibinfo {year}
  {2016})}\BibitemShut {NoStop}%
\bibitem [{\citenamefont {{Hasegawa}}\ \emph {et~al.}(2014)\citenamefont
  {{Hasegawa}}, \citenamefont {{Sonnerup}}, \citenamefont {{Hu}},\ and\
  \citenamefont {{Nakamura}}}]{HasegawaEA14}%
  \BibitemOpen
  \bibfield  {author} {\bibinfo {author} {\bibfnamefont {H.}~\bibnamefont
  {{Hasegawa}}}, \bibinfo {author} {\bibfnamefont {B.~U.~{\"O}.}\ \bibnamefont
  {{Sonnerup}}}, \bibinfo {author} {\bibfnamefont {Q.}~\bibnamefont {{Hu}}},\
  and\ \bibinfo {author} {\bibfnamefont {T.}~\bibnamefont {{Nakamura}}},\
  }\bibfield  {title} {\bibinfo {title} {{Reconstruction of an evolving
  magnetic flux rope in the solar wind: Decomposing spatial and temporal
  variations from single-spacecraft data}},\ }\href
  {https://doi.org/10.1002/2013JA019180} {\bibfield  {journal} {\bibinfo
  {journal} {Journal of Geophysical Research (Space Physics)}\ }\textbf
  {\bibinfo {volume} {119}},\ \bibinfo {pages} {97} (\bibinfo {year}
  {2014})}\BibitemShut {NoStop}%
\bibitem [{\citenamefont {Montgomery}\ \emph {et~al.}(1979)\citenamefont
  {Montgomery}, \citenamefont {Turner},\ and\ \citenamefont
  {Vahala}}]{MontEA79}%
  \BibitemOpen
  \bibfield  {author} {\bibinfo {author} {\bibfnamefont {D.}~\bibnamefont
  {Montgomery}}, \bibinfo {author} {\bibfnamefont {L.}~\bibnamefont {Turner}},\
  and\ \bibinfo {author} {\bibfnamefont {G.}~\bibnamefont {Vahala}},\
  }\bibfield  {title} {\bibinfo {title} {Most probable states in
  magnetohydrodynamics},\ }\href {https://doi.org/10.1017/S0022377800021802}
  {\bibfield  {journal} {\bibinfo  {journal} {J.\ Plasma Phys.}\ }\textbf
  {\bibinfo {volume} {21}},\ \bibinfo {pages} {239} (\bibinfo {year}
  {1979})}\BibitemShut {NoStop}%
\bibitem [{\citenamefont {Osman}\ \emph {et~al.}(2011)\citenamefont {Osman},
  \citenamefont {Wan}, \citenamefont {Matthaeus}, \citenamefont {Breech},\ and\
  \citenamefont {Oughton}}]{osman2011directional}%
  \BibitemOpen
  \bibfield  {author} {\bibinfo {author} {\bibfnamefont {K.~T.}\ \bibnamefont
  {Osman}}, \bibinfo {author} {\bibfnamefont {M.}~\bibnamefont {Wan}}, \bibinfo
  {author} {\bibfnamefont {W.~H.}\ \bibnamefont {Matthaeus}}, \bibinfo {author}
  {\bibfnamefont {B.}~\bibnamefont {Breech}},\ and\ \bibinfo {author}
  {\bibfnamefont {S.}~\bibnamefont {Oughton}},\ }\bibfield  {title} {\bibinfo
  {title} {Directional alignment and non-gaussian statistics in solar wind
  turbulence},\ }\href {https://doi.org/10.1088/0004-637x/741/2/75} {\bibfield
  {journal} {\bibinfo  {journal} {The Astrophysical Journal}\ }\textbf
  {\bibinfo {volume} {741}},\ \bibinfo {pages} {75} (\bibinfo {year}
  {2011})}\BibitemShut {NoStop}%
\bibitem [{\citenamefont {Matthaeus}\ \emph {et~al.}(2015)\citenamefont
  {Matthaeus}, \citenamefont {Wan}, \citenamefont {Servidio}, \citenamefont
  {Greco}, \citenamefont {Osman}, \citenamefont {Oughton},\ and\ \citenamefont
  {Dmitruk}}]{matthaeus2015intermittency}%
  \BibitemOpen
  \bibfield  {author} {\bibinfo {author} {\bibfnamefont {W.~H.}\ \bibnamefont
  {Matthaeus}}, \bibinfo {author} {\bibfnamefont {M.}~\bibnamefont {Wan}},
  \bibinfo {author} {\bibfnamefont {S.}~\bibnamefont {Servidio}}, \bibinfo
  {author} {\bibfnamefont {A.}~\bibnamefont {Greco}}, \bibinfo {author}
  {\bibfnamefont {K.~T.}\ \bibnamefont {Osman}}, \bibinfo {author}
  {\bibfnamefont {S.}~\bibnamefont {Oughton}},\ and\ \bibinfo {author}
  {\bibfnamefont {P.}~\bibnamefont {Dmitruk}},\ }\bibfield  {title} {\bibinfo
  {title} {Intermittency, nonlinear dynamics and dissipation in the solar wind
  and astrophysical plasmas},\ }\href {https://doi.org/10.1098/rsta.2014.0154}
  {\bibfield  {journal} {\bibinfo  {journal} {Philosophical Transactions of the
  Royal Society A: Mathematical, Physical and Engineering Sciences}\ }\textbf
  {\bibinfo {volume} {373}},\ \bibinfo {pages} {20140154} (\bibinfo {year}
  {2015})}\BibitemShut {NoStop}%
\bibitem [{\citenamefont {Servidio}\ \emph {et~al.}(2014)\citenamefont
  {Servidio}, \citenamefont {Gurgiolo}, \citenamefont {Carbone},\ and\
  \citenamefont {Goldstein}}]{servidio2014relaxation}%
  \BibitemOpen
  \bibfield  {author} {\bibinfo {author} {\bibfnamefont {S.}~\bibnamefont
  {Servidio}}, \bibinfo {author} {\bibfnamefont {C.}~\bibnamefont {Gurgiolo}},
  \bibinfo {author} {\bibfnamefont {V.}~\bibnamefont {Carbone}},\ and\ \bibinfo
  {author} {\bibfnamefont {M.~L.}\ \bibnamefont {Goldstein}},\ }\bibfield
  {title} {\bibinfo {title} {Relaxation processes in solar wind turbulence},\
  }\href {https://doi.org/10.1088/2041-8205/789/2/l44} {\bibfield  {journal}
  {\bibinfo  {journal} {The Astrophysical Journal}\ }\textbf {\bibinfo {volume}
  {789}},\ \bibinfo {pages} {L44} (\bibinfo {year} {2014})}\BibitemShut
  {NoStop}%
\bibitem [{\citenamefont {Burch}\ \emph {et~al.}(2016)\citenamefont {Burch},
  \citenamefont {Moore}, \citenamefont {Torbert},\ and\ \citenamefont
  {Giles}}]{burch2016magnetospheric}%
  \BibitemOpen
  \bibfield  {author} {\bibinfo {author} {\bibfnamefont {J.~L.}\ \bibnamefont
  {Burch}}, \bibinfo {author} {\bibfnamefont {T.~E.}\ \bibnamefont {Moore}},
  \bibinfo {author} {\bibfnamefont {R.~B.}\ \bibnamefont {Torbert}},\ and\
  \bibinfo {author} {\bibfnamefont {B.~L.}\ \bibnamefont {Giles}},\ }\bibfield
  {title} {\bibinfo {title} {Magnetospheric multiscale overview and science
  objectives},\ }\href {https://doi.org/10.1007/s11214-015-0164-9} {\bibfield
  {journal} {\bibinfo  {journal} {Space Science Reviews}\ }\textbf {\bibinfo
  {volume} {199}},\ \bibinfo {pages} {5} (\bibinfo {year} {2016})}\BibitemShut
  {NoStop}%
\bibitem [{\citenamefont {Matthaeus}\ \emph {et~al.}(1998)\citenamefont
  {Matthaeus}, \citenamefont {Smith},\ and\ \citenamefont
  {Oughton}}]{matthaeus1998dynamical}%
  \BibitemOpen
  \bibfield  {author} {\bibinfo {author} {\bibfnamefont {W.~H.}\ \bibnamefont
  {Matthaeus}}, \bibinfo {author} {\bibfnamefont {C.~W.}\ \bibnamefont
  {Smith}},\ and\ \bibinfo {author} {\bibfnamefont {S.}~\bibnamefont
  {Oughton}},\ }\bibfield  {title} {\bibinfo {title} {Dynamical age of solar
  wind turbulence in the outer heliosphere},\ }\href
  {https://doi.org/https://doi.org/10.1029/97JA03729} {\bibfield  {journal}
  {\bibinfo  {journal} {Journal of Geophysical Research: Space Physics}\
  }\textbf {\bibinfo {volume} {103}},\ \bibinfo {pages} {6495} (\bibinfo {year}
  {1998})}\BibitemShut {NoStop}%
\bibitem [{\citenamefont {{Matthaeus}}\ \emph {et~al.}(1982)\citenamefont
  {{Matthaeus}}, \citenamefont {{Goldstein}},\ and\ \citenamefont
  {{Smith}}}]{matthaeus1982evaluation}%
  \BibitemOpen
  \bibfield  {author} {\bibinfo {author} {\bibfnamefont {W.~H.}\ \bibnamefont
  {{Matthaeus}}}, \bibinfo {author} {\bibfnamefont {M.~L.}\ \bibnamefont
  {{Goldstein}}},\ and\ \bibinfo {author} {\bibfnamefont {C.}~\bibnamefont
  {{Smith}}},\ }\bibfield  {title} {\bibinfo {title} {{Evaluation of magnetic
  helicity in homogeneous turbulence}},\ }\href
  {https://doi.org/10.1103/PhysRevLett.48.1256} {\bibfield  {journal} {\bibinfo
   {journal} {\prl}\ }\textbf {\bibinfo {volume} {48}},\ \bibinfo {pages}
  {1256} (\bibinfo {year} {1982})}\BibitemShut {NoStop}%
\bibitem [{\citenamefont {Woltjer}(1958{\natexlab{b}})}]{woltjer1958theorem}%
  \BibitemOpen
  \bibfield  {author} {\bibinfo {author} {\bibfnamefont {L.}~\bibnamefont
  {Woltjer}},\ }\bibfield  {title} {\bibinfo {title} {A theorem on force-free
  magnetic fields},\ }\href {https://doi.org/10.1073/pnas.44.6.489} {\bibfield
  {journal} {\bibinfo  {journal} {Proceedings of the National Academy of
  Sciences}\ }\textbf {\bibinfo {volume} {44}},\ \bibinfo {pages} {489}
  (\bibinfo {year} {1958}{\natexlab{b}})}\BibitemShut {NoStop}%
\bibitem [{\citenamefont {Pollock}\ \emph {et~al.}(2016)\citenamefont
  {Pollock}, \citenamefont {Moore}, \citenamefont {Jacques}, \citenamefont
  {Burch}, \citenamefont {Gliese}, \citenamefont {Saito}, \citenamefont
  {Omoto}, \citenamefont {Avanov}, \citenamefont {Barrie}, \citenamefont
  {Coffey}, \citenamefont {Dorelli}, \citenamefont {Gershman}, \citenamefont
  {Giles}, \citenamefont {Rosnack}, \citenamefont {Salo}, \citenamefont
  {Yokota}, \citenamefont {Adrian}, \citenamefont {Aoustin}, \citenamefont
  {Auletti}, \citenamefont {Aung}, \citenamefont {Bigio}, \citenamefont {Cao},
  \citenamefont {Chandler}, \citenamefont {Chornay}, \citenamefont {Christian},
  \citenamefont {Clark}, \citenamefont {Collinson}, \citenamefont {Corris},
  \citenamefont {De Los Santos}, \citenamefont {Devlin}, \citenamefont
  {Diaz}, \citenamefont {Dickerson}, \citenamefont {Dickson}, \citenamefont
  {Diekmann}, \citenamefont {Diggs}, \citenamefont {Duncan}, \citenamefont
  {Figueroa-Vinas}, \citenamefont {Firman}, \citenamefont {Freeman},
  \citenamefont {Galassi}, \citenamefont {Garcia}, \citenamefont {Goodhart},
  \citenamefont {Guererro}, \citenamefont {Hageman}, \citenamefont {Hanley},
  \citenamefont {Hemminger}, \citenamefont {Holland}, \citenamefont {Hutchins},
  \citenamefont {James}, \citenamefont {Jones}, \citenamefont {Kreisler},
  \citenamefont {Kujawski}, \citenamefont {Lavu}, \citenamefont {Lobell},
  \citenamefont {LeCompte}, \citenamefont {Lukemire}, \citenamefont
  {MacDonald}, \citenamefont {Mariano}, \citenamefont {Mukai}, \citenamefont
  {Narayanan}, \citenamefont {Nguyan}, \citenamefont {Onizuka}, \citenamefont
  {Paterson}, \citenamefont {Persyn}, \citenamefont {Piepgrass}, \citenamefont
  {Cheney}, \citenamefont {Rager}, \citenamefont {Raghuram}, \citenamefont
  {Ramil}, \citenamefont {Reichenthal}, \citenamefont {Rodriguez},
  \citenamefont {Rouzaud}, \citenamefont {Rucker}, \citenamefont {Samara},
  \citenamefont {Sauvaud}, \citenamefont {Schuster}, \citenamefont {Shappirio},
  \citenamefont {Shelton}, \citenamefont {Sher}, \citenamefont {Smith},
  \citenamefont {Smith}, \citenamefont {Smith}, \citenamefont {Steinfeld},
  \citenamefont {Szymkiewicz}, \citenamefont {Tanimoto}, \citenamefont
  {Taylor}, \citenamefont {Tucker}, \citenamefont {Tull}, \citenamefont {Uhl},
  \citenamefont {Vloet}, \citenamefont {Walpole}, \citenamefont {Weidner},
  \citenamefont {White}, \citenamefont {Winkert}, \citenamefont {Yeh},\ and\
  \citenamefont {Zeuch}}]{pollock2016fast}%
  \BibitemOpen
  \bibfield  {author} {\bibinfo {author} {\bibfnamefont {C.}~\bibnamefont
  {Pollock}}, \bibinfo {author} {\bibfnamefont {T.}~\bibnamefont {Moore}},
  \bibinfo {author} {\bibfnamefont {A.}~\bibnamefont {Jacques}}, \bibinfo
  {author} {\bibfnamefont {J.}~\bibnamefont {Burch}}, \bibinfo {author}
  {\bibfnamefont {U.}~\bibnamefont {Gliese}}, \bibinfo {author} {\bibfnamefont
  {Y.}~\bibnamefont {Saito}}, \bibinfo {author} {\bibfnamefont
  {T.}~\bibnamefont {Omoto}}, \bibinfo {author} {\bibfnamefont
  {L.}~\bibnamefont {Avanov}}, \bibinfo {author} {\bibfnamefont
  {A.}~\bibnamefont {Barrie}}, \bibinfo {author} {\bibfnamefont
  {V.}~\bibnamefont {Coffey}}, \bibinfo {author} {\bibfnamefont
  {J.}~\bibnamefont {Dorelli}}, \bibinfo {author} {\bibfnamefont
  {D.}~\bibnamefont {Gershman}}, \bibinfo {author} {\bibfnamefont
  {B.}~\bibnamefont {Giles}}, \bibinfo {author} {\bibfnamefont
  {T.}~\bibnamefont {Rosnack}}, \bibinfo {author} {\bibfnamefont
  {C.}~\bibnamefont {Salo}}, \bibinfo {author} {\bibfnamefont {S.}~\bibnamefont
  {Yokota}}, \bibinfo {author} {\bibfnamefont {M.}~\bibnamefont {Adrian}},
  \bibinfo {author} {\bibfnamefont {C.}~\bibnamefont {Aoustin}}, \bibinfo
  {author} {\bibfnamefont {C.}~\bibnamefont {Auletti}}, \bibinfo {author}
  {\bibfnamefont {S.}~\bibnamefont {Aung}}, \bibinfo {author} {\bibfnamefont
  {V.}~\bibnamefont {Bigio}}, \bibinfo {author} {\bibfnamefont
  {N.}~\bibnamefont {Cao}}, \bibinfo {author} {\bibfnamefont {M.}~\bibnamefont
  {Chandler}}, \bibinfo {author} {\bibfnamefont {D.}~\bibnamefont {Chornay}},
  \bibinfo {author} {\bibfnamefont {K.}~\bibnamefont {Christian}}, \bibinfo
  {author} {\bibfnamefont {G.}~\bibnamefont {Clark}}, \bibinfo {author}
  {\bibfnamefont {G.}~\bibnamefont {Collinson}}, \bibinfo {author}
  {\bibfnamefont {T.}~\bibnamefont {Corris}}, \bibinfo {author} {\bibfnamefont
  {A.}~\bibnamefont {De Los Santos}}, \bibinfo {author} {\bibfnamefont
  {R.}~\bibnamefont {Devlin}}, \bibinfo {author} {\bibfnamefont
  {T.}~\bibnamefont {Diaz}}, \bibinfo {author} {\bibfnamefont {T.}~\bibnamefont
  {Dickerson}}, \bibinfo {author} {\bibfnamefont {C.}~\bibnamefont {Dickson}},
  \bibinfo {author} {\bibfnamefont {A.}~\bibnamefont {Diekmann}}, \bibinfo
  {author} {\bibfnamefont {F.}~\bibnamefont {Diggs}}, \bibinfo {author}
  {\bibfnamefont {C.}~\bibnamefont {Duncan}}, \bibinfo {author} {\bibfnamefont
  {A.}~\bibnamefont {Figueroa-Vinas}}, \bibinfo {author} {\bibfnamefont
  {C.}~\bibnamefont {Firman}}, \bibinfo {author} {\bibfnamefont
  {M.}~\bibnamefont {Freeman}}, \bibinfo {author} {\bibfnamefont
  {N.}~\bibnamefont {Galassi}}, \bibinfo {author} {\bibfnamefont
  {K.}~\bibnamefont {Garcia}}, \bibinfo {author} {\bibfnamefont
  {G.}~\bibnamefont {Goodhart}}, \bibinfo {author} {\bibfnamefont
  {D.}~\bibnamefont {Guererro}}, \bibinfo {author} {\bibfnamefont
  {J.}~\bibnamefont {Hageman}}, \bibinfo {author} {\bibfnamefont
  {J.}~\bibnamefont {Hanley}}, \bibinfo {author} {\bibfnamefont
  {E.}~\bibnamefont {Hemminger}}, \bibinfo {author} {\bibfnamefont
  {M.}~\bibnamefont {Holland}}, \bibinfo {author} {\bibfnamefont
  {M.}~\bibnamefont {Hutchins}}, \bibinfo {author} {\bibfnamefont
  {T.}~\bibnamefont {James}}, \bibinfo {author} {\bibfnamefont
  {W.}~\bibnamefont {Jones}}, \bibinfo {author} {\bibfnamefont
  {S.}~\bibnamefont {Kreisler}}, \bibinfo {author} {\bibfnamefont
  {J.}~\bibnamefont {Kujawski}}, \bibinfo {author} {\bibfnamefont
  {V.}~\bibnamefont {Lavu}}, \bibinfo {author} {\bibfnamefont {J.}~\bibnamefont
  {Lobell}}, \bibinfo {author} {\bibfnamefont {E.}~\bibnamefont {LeCompte}},
  \bibinfo {author} {\bibfnamefont {A.}~\bibnamefont {Lukemire}}, \bibinfo
  {author} {\bibfnamefont {E.}~\bibnamefont {MacDonald}}, \bibinfo {author}
  {\bibfnamefont {A.}~\bibnamefont {Mariano}}, \bibinfo {author} {\bibfnamefont
  {T.}~\bibnamefont {Mukai}}, \bibinfo {author} {\bibfnamefont
  {K.}~\bibnamefont {Narayanan}}, \bibinfo {author} {\bibfnamefont
  {Q.}~\bibnamefont {Nguyan}}, \bibinfo {author} {\bibfnamefont
  {M.}~\bibnamefont {Onizuka}}, \bibinfo {author} {\bibfnamefont
  {W.}~\bibnamefont {Paterson}}, \bibinfo {author} {\bibfnamefont
  {S.}~\bibnamefont {Persyn}}, \bibinfo {author} {\bibfnamefont
  {B.}~\bibnamefont {Piepgrass}}, \bibinfo {author} {\bibfnamefont
  {F.}~\bibnamefont {Cheney}}, \bibinfo {author} {\bibfnamefont
  {A.}~\bibnamefont {Rager}}, \bibinfo {author} {\bibfnamefont
  {T.}~\bibnamefont {Raghuram}}, \bibinfo {author} {\bibfnamefont
  {A.}~\bibnamefont {Ramil}}, \bibinfo {author} {\bibfnamefont
  {L.}~\bibnamefont {Reichenthal}}, \bibinfo {author} {\bibfnamefont
  {H.}~\bibnamefont {Rodriguez}}, \bibinfo {author} {\bibfnamefont
  {J.}~\bibnamefont {Rouzaud}}, \bibinfo {author} {\bibfnamefont
  {A.}~\bibnamefont {Rucker}}, \bibinfo {author} {\bibfnamefont
  {M.}~\bibnamefont {Samara}}, \bibinfo {author} {\bibfnamefont {J.-A.}\
  \bibnamefont {Sauvaud}}, \bibinfo {author} {\bibfnamefont {D.}~\bibnamefont
  {Schuster}}, \bibinfo {author} {\bibfnamefont {M.}~\bibnamefont {Shappirio}},
  \bibinfo {author} {\bibfnamefont {K.}~\bibnamefont {Shelton}}, \bibinfo
  {author} {\bibfnamefont {D.}~\bibnamefont {Sher}}, \bibinfo {author}
  {\bibfnamefont {D.}~\bibnamefont {Smith}}, \bibinfo {author} {\bibfnamefont
  {K.}~\bibnamefont {Smith}}, \bibinfo {author} {\bibfnamefont
  {S.}~\bibnamefont {Smith}}, \bibinfo {author} {\bibfnamefont
  {D.}~\bibnamefont {Steinfeld}}, \bibinfo {author} {\bibfnamefont
  {R.}~\bibnamefont {Szymkiewicz}}, \bibinfo {author} {\bibfnamefont
  {K.}~\bibnamefont {Tanimoto}}, \bibinfo {author} {\bibfnamefont
  {J.}~\bibnamefont {Taylor}}, \bibinfo {author} {\bibfnamefont
  {C.}~\bibnamefont {Tucker}}, \bibinfo {author} {\bibfnamefont
  {K.}~\bibnamefont {Tull}}, \bibinfo {author} {\bibfnamefont {A.}~\bibnamefont
  {Uhl}}, \bibinfo {author} {\bibfnamefont {J.}~\bibnamefont {Vloet}}, \bibinfo
  {author} {\bibfnamefont {P.}~\bibnamefont {Walpole}}, \bibinfo {author}
  {\bibfnamefont {S.}~\bibnamefont {Weidner}}, \bibinfo {author} {\bibfnamefont
  {D.}~\bibnamefont {White}}, \bibinfo {author} {\bibfnamefont
  {G.}~\bibnamefont {Winkert}}, \bibinfo {author} {\bibfnamefont {P.-S.}\
  \bibnamefont {Yeh}},\ and\ \bibinfo {author} {\bibfnamefont {M.}~\bibnamefont
  {Zeuch}},\ }\bibfield  {title} {\bibinfo {title} {Fast plasma investigation
  for magnetospheric multiscale},\ }\href
  {https://doi.org/10.1007/s11214-016-0245-4} {\bibfield  {journal} {\bibinfo
  {journal} {Space Science Reviews}\ }\textbf {\bibinfo {volume} {199}},\
  \bibinfo {pages} {331} (\bibinfo {year} {2016})}\BibitemShut {NoStop}%
\bibitem [{\citenamefont {Russell}\ \emph {et~al.}(2016)\citenamefont
  {Russell}, \citenamefont {Anderson}, \citenamefont {Baumjohann},
  \citenamefont {Bromund}, \citenamefont {Dearborn}, \citenamefont {Fischer},
  \citenamefont {Le}, \citenamefont {Leinweber}, \citenamefont {Leneman},
  \citenamefont {Magnes}, \citenamefont {Means}, \citenamefont {Moldwin},
  \citenamefont {Nakamura}, \citenamefont {Pierce}, \citenamefont {Plaschke},
  \citenamefont {Rowe}, \citenamefont {Slavin}, \citenamefont {Strangeway},
  \citenamefont {Torbert}, \citenamefont {Hagen}, \citenamefont {Jernej},
  \citenamefont {Valavanoglou},\ and\ \citenamefont
  {Richter}}]{russell2016magnetospheric}%
  \BibitemOpen
  \bibfield  {author} {\bibinfo {author} {\bibfnamefont {C.~T.}\ \bibnamefont
  {Russell}}, \bibinfo {author} {\bibfnamefont {B.~J.}\ \bibnamefont
  {Anderson}}, \bibinfo {author} {\bibfnamefont {W.}~\bibnamefont
  {Baumjohann}}, \bibinfo {author} {\bibfnamefont {K.~R.}\ \bibnamefont
  {Bromund}}, \bibinfo {author} {\bibfnamefont {D.}~\bibnamefont {Dearborn}},
  \bibinfo {author} {\bibfnamefont {D.}~\bibnamefont {Fischer}}, \bibinfo
  {author} {\bibfnamefont {G.}~\bibnamefont {Le}}, \bibinfo {author}
  {\bibfnamefont {H.~K.}\ \bibnamefont {Leinweber}}, \bibinfo {author}
  {\bibfnamefont {D.}~\bibnamefont {Leneman}}, \bibinfo {author} {\bibfnamefont
  {W.}~\bibnamefont {Magnes}}, \bibinfo {author} {\bibfnamefont {J.~D.}\
  \bibnamefont {Means}}, \bibinfo {author} {\bibfnamefont {M.~B.}\ \bibnamefont
  {Moldwin}}, \bibinfo {author} {\bibfnamefont {R.}~\bibnamefont {Nakamura}},
  \bibinfo {author} {\bibfnamefont {D.}~\bibnamefont {Pierce}}, \bibinfo
  {author} {\bibfnamefont {F.}~\bibnamefont {Plaschke}}, \bibinfo {author}
  {\bibfnamefont {K.~M.}\ \bibnamefont {Rowe}}, \bibinfo {author}
  {\bibfnamefont {J.~A.}\ \bibnamefont {Slavin}}, \bibinfo {author}
  {\bibfnamefont {R.~J.}\ \bibnamefont {Strangeway}}, \bibinfo {author}
  {\bibfnamefont {R.}~\bibnamefont {Torbert}}, \bibinfo {author} {\bibfnamefont
  {C.}~\bibnamefont {Hagen}}, \bibinfo {author} {\bibfnamefont
  {I.}~\bibnamefont {Jernej}}, \bibinfo {author} {\bibfnamefont
  {A.}~\bibnamefont {Valavanoglou}},\ and\ \bibinfo {author} {\bibfnamefont
  {I.}~\bibnamefont {Richter}},\ }\bibfield  {title} {\bibinfo {title} {The
  magnetospheric multiscale magnetometers},\ }\href
  {https://doi.org/10.1007/s11214-014-0057-3} {\bibfield  {journal} {\bibinfo
  {journal} {Space Science Reviews}\ }\textbf {\bibinfo {volume} {199}},\
  \bibinfo {pages} {189} (\bibinfo {year} {2016})}\BibitemShut {NoStop}%
\bibitem [{\citenamefont {Shue}\ \emph {et~al.}(1997)\citenamefont {Shue},
  \citenamefont {Chao}, \citenamefont {Fu}, \citenamefont {Russell},
  \citenamefont {Song}, \citenamefont {Khurana},\ and\ \citenamefont
  {Singer}}]{shue1997new}%
  \BibitemOpen
  \bibfield  {author} {\bibinfo {author} {\bibfnamefont {J.-H.}\ \bibnamefont
  {Shue}}, \bibinfo {author} {\bibfnamefont {J.~K.}\ \bibnamefont {Chao}},
  \bibinfo {author} {\bibfnamefont {H.~C.}\ \bibnamefont {Fu}}, \bibinfo
  {author} {\bibfnamefont {C.~T.}\ \bibnamefont {Russell}}, \bibinfo {author}
  {\bibfnamefont {P.}~\bibnamefont {Song}}, \bibinfo {author} {\bibfnamefont
  {K.~K.}\ \bibnamefont {Khurana}},\ and\ \bibinfo {author} {\bibfnamefont
  {H.~J.}\ \bibnamefont {Singer}},\ }\bibfield  {title} {\bibinfo {title} {A
  new functional form to study the solar wind control of the magnetopause size
  and shape},\ }\href {https://doi.org/https://doi.org/10.1029/97JA00196}
  {\bibfield  {journal} {\bibinfo  {journal} {Journal of Geophysical Research:
  Space Physics}\ }\textbf {\bibinfo {volume} {102}},\ \bibinfo {pages} {9497}
  (\bibinfo {year} {1997})},\ \Eprint
  {https://arxiv.org/abs/https://agupubs.onlinelibrary.wiley.com/doi/pdf/10.1029/97JA00196}
  {https://agupubs.onlinelibrary.wiley.com/doi/pdf/10.1029/97JA00196}
  \BibitemShut {NoStop}%
\bibitem [{\citenamefont {Chanteur}(1998)}]{chanteur1998spatial}%
  \BibitemOpen
  \bibfield  {author} {\bibinfo {author} {\bibfnamefont {G.}~\bibnamefont
  {Chanteur}},\ }\bibfield  {title} {\bibinfo {title} {Spatial interpolation
  for four spacecraft: Theory},\ }\href@noop {} {\bibfield  {journal} {\bibinfo
   {journal} {ISSI Scientific Reports Series}\ }\textbf {\bibinfo {volume}
  {1}},\ \bibinfo {pages} {349} (\bibinfo {year} {1998})}\BibitemShut {NoStop}%
\bibitem [{\citenamefont {{Chhiber}}\ \emph {et~al.}(2018)\citenamefont
  {{Chhiber}}, \citenamefont {{Chasapis}}, \citenamefont {{Bandyopadhyay}},
  \citenamefont {{Parashar}}, \citenamefont {{Matthaeus}}, \citenamefont
  {{Maruca}}, \citenamefont {{Moore}}, \citenamefont {{Burch}}, \citenamefont
  {{Torbert}}, \citenamefont {{Russell}}, \citenamefont {{Le Contel}},
  \citenamefont {{Argall}}, \citenamefont {{Fischer}}, \citenamefont
  {{Mirioni}}, \citenamefont {{Strangeway}}, \citenamefont {{Pollock}},
  \citenamefont {{Giles}},\ and\ \citenamefont
  {{Gershman}}}]{chhiber2018higher}%
  \BibitemOpen
  \bibfield  {author} {\bibinfo {author} {\bibfnamefont {R.}~\bibnamefont
  {{Chhiber}}}, \bibinfo {author} {\bibfnamefont {A.}~\bibnamefont
  {{Chasapis}}}, \bibinfo {author} {\bibfnamefont {R.}~\bibnamefont
  {{Bandyopadhyay}}}, \bibinfo {author} {\bibfnamefont {T.~N.}\ \bibnamefont
  {{Parashar}}}, \bibinfo {author} {\bibfnamefont {W.~H.}\ \bibnamefont
  {{Matthaeus}}}, \bibinfo {author} {\bibfnamefont {B.~A.}\ \bibnamefont
  {{Maruca}}}, \bibinfo {author} {\bibfnamefont {T.~E.}\ \bibnamefont
  {{Moore}}}, \bibinfo {author} {\bibfnamefont {J.~L.}\ \bibnamefont
  {{Burch}}}, \bibinfo {author} {\bibfnamefont {R.~B.}\ \bibnamefont
  {{Torbert}}}, \bibinfo {author} {\bibfnamefont {C.~T.}\ \bibnamefont
  {{Russell}}}, \bibinfo {author} {\bibfnamefont {O.}~\bibnamefont {{Le
  Contel}}}, \bibinfo {author} {\bibfnamefont {M.~R.}\ \bibnamefont
  {{Argall}}}, \bibinfo {author} {\bibfnamefont {D.}~\bibnamefont {{Fischer}}},
  \bibinfo {author} {\bibfnamefont {L.}~\bibnamefont {{Mirioni}}}, \bibinfo
  {author} {\bibfnamefont {R.~J.}\ \bibnamefont {{Strangeway}}}, \bibinfo
  {author} {\bibfnamefont {C.~J.}\ \bibnamefont {{Pollock}}}, \bibinfo {author}
  {\bibfnamefont {B.~L.}\ \bibnamefont {{Giles}}},\ and\ \bibinfo {author}
  {\bibfnamefont {D.~J.}\ \bibnamefont {{Gershman}}},\ }\bibfield  {title}
  {\bibinfo {title} {{Higher-Order Turbulence Statistics in the Earth's
  Magnetosheath and the Solar Wind Using Magnetospheric Multiscale
  Observations}},\ }\href {https://doi.org/10.1029/2018JA025768} {\bibfield
  {journal} {\bibinfo  {journal} {Journal of Geophysical Research (Space
  Physics)}\ }\textbf {\bibinfo {volume} {123}},\ \bibinfo {pages} {9941}
  (\bibinfo {year} {2018})}\BibitemShut {NoStop}%
\bibitem [{\citenamefont {Yordanova}\ \emph {et~al.}(2011)\citenamefont
  {Yordanova}, \citenamefont {Perri},\ and\ \citenamefont
  {Carbone}}]{yordanova2011reduced}%
  \BibitemOpen
  \bibfield  {author} {\bibinfo {author} {\bibfnamefont {E.}~\bibnamefont
  {Yordanova}}, \bibinfo {author} {\bibfnamefont {S.}~\bibnamefont {Perri}},\
  and\ \bibinfo {author} {\bibfnamefont {V.}~\bibnamefont {Carbone}},\
  }\bibfield  {title} {\bibinfo {title} {Reduced magnetic helicity behavior in
  different plasma regions of near-earth space},\ }\href
  {https://doi.org/https://doi.org/10.1029/2010JA015875} {\bibfield  {journal}
  {\bibinfo  {journal} {Journal of Geophysical Research: Space Physics}\
  }\textbf {\bibinfo {volume} {116}} (\bibinfo {year} {2011})}\BibitemShut
  {NoStop}%
\bibitem [{\citenamefont {Pecora}\ \emph {et~al.}(2021)\citenamefont {Pecora},
  \citenamefont {Servidio}, \citenamefont {Greco},\ and\ \citenamefont
  {Matthaeus}}]{pecora2021identification}%
  \BibitemOpen
  \bibfield  {author} {\bibinfo {author} {\bibfnamefont {F.}~\bibnamefont
  {Pecora}}, \bibinfo {author} {\bibfnamefont {S.}~\bibnamefont {Servidio}},
  \bibinfo {author} {\bibfnamefont {A.}~\bibnamefont {Greco}},\ and\ \bibinfo
  {author} {\bibfnamefont {W.~H.}\ \bibnamefont {Matthaeus}},\ }\bibfield
  {title} {\bibinfo {title} {Identification of coherent structures in space
  plasmas: the magnetic helicity-pvi method},\ }\href
  {https://doi.org/10.1051/0004-6361/202039639} {\bibfield  {journal} {\bibinfo
   {journal} {A\&A}\ }\textbf {\bibinfo {volume} {650}},\ \bibinfo {pages}
  {A20} (\bibinfo {year} {2021})}\BibitemShut {NoStop}%
\bibitem [{\citenamefont {Burlaga}(1988)}]{Burlaga88}%
  \BibitemOpen
  \bibfield  {author} {\bibinfo {author} {\bibfnamefont {L.}~\bibnamefont
  {Burlaga}},\ }\bibfield  {title} {\bibinfo {title} {Magnetic clouds and
  force-free fields with constant alpha},\ }\href@noop {} {\bibfield  {journal}
  {\bibinfo  {journal} {Journal of Geophysical Research: Space Physics}\
  }\textbf {\bibinfo {volume} {93}},\ \bibinfo {pages} {7217} (\bibinfo {year}
  {1988})}\BibitemShut {NoStop}%
\bibitem [{\citenamefont {Pouquet}(1996)}]{pouquet1996turbulence}%
  \BibitemOpen
  \bibfield  {author} {\bibinfo {author} {\bibfnamefont {A.}~\bibnamefont
  {Pouquet}},\ }\bibfield  {title} {\bibinfo {title} {Turbulence, statistics
  and structures: an introduction},\ }in\ \href@noop {} {\emph {\bibinfo
  {booktitle} {Plasma Astrophysics}}},\ \bibinfo {editor} {edited by\ \bibinfo
  {editor} {\bibfnamefont {C.}~\bibnamefont {Chiuderi}}\ and\ \bibinfo {editor}
  {\bibfnamefont {G.}~\bibnamefont {Einaudi}}}\ (\bibinfo  {publisher}
  {Springer Berlin Heidelberg},\ \bibinfo {address} {Berlin, Heidelberg},\
  \bibinfo {year} {1996})\ pp.\ \bibinfo {pages} {163--212}\BibitemShut
  {NoStop}%
\bibitem [{\citenamefont {Zank}\ \emph {et~al.}(2021)\citenamefont {Zank},
  \citenamefont {Nakanotani}, \citenamefont {Zhao}, \citenamefont {Du},
  \citenamefont {Adhikari}, \citenamefont {Che},\ and\ \citenamefont
  {le~Roux}}]{ZankEA21-shocks}%
  \BibitemOpen
  \bibfield  {author} {\bibinfo {author} {\bibfnamefont {G.~P.}\ \bibnamefont
  {Zank}}, \bibinfo {author} {\bibfnamefont {M.}~\bibnamefont {Nakanotani}},
  \bibinfo {author} {\bibfnamefont {L.~L.}\ \bibnamefont {Zhao}}, \bibinfo
  {author} {\bibfnamefont {S.}~\bibnamefont {Du}}, \bibinfo {author}
  {\bibfnamefont {L.}~\bibnamefont {Adhikari}}, \bibinfo {author}
  {\bibfnamefont {H.}~\bibnamefont {Che}},\ and\ \bibinfo {author}
  {\bibfnamefont {J.~A.}\ \bibnamefont {le~Roux}},\ }\bibfield  {title}
  {\bibinfo {title} {Flux ropes, turbulence, and collisionless perpendicular
  shock waves: High plasma beta case},\ }\href
  {https://doi.org/10.3847/1538-4357/abf7c8} {\bibfield  {journal} {\bibinfo
  {journal} {The Astrophysical Journal}\ }\textbf {\bibinfo {volume} {913}},\
  \bibinfo {pages} {127} (\bibinfo {year} {2021})}\BibitemShut {NoStop}%
\bibitem [{\citenamefont {Trotta}\ \emph {et~al.}(2022)\citenamefont {Trotta},
  \citenamefont {Pecora}, \citenamefont {Settino}, \citenamefont {Perrone},
  \citenamefont {Hietala}, \citenamefont {Horbury}, \citenamefont {Matthaeus},
  \citenamefont {Burgess}, \citenamefont {Servidio},\ and\ \citenamefont
  {Valentini}}]{trotta2022transmission}%
  \BibitemOpen
  \bibfield  {author} {\bibinfo {author} {\bibfnamefont {D.}~\bibnamefont
  {Trotta}}, \bibinfo {author} {\bibfnamefont {F.}~\bibnamefont {Pecora}},
  \bibinfo {author} {\bibfnamefont {A.}~\bibnamefont {Settino}}, \bibinfo
  {author} {\bibfnamefont {D.}~\bibnamefont {Perrone}}, \bibinfo {author}
  {\bibfnamefont {H.}~\bibnamefont {Hietala}}, \bibinfo {author} {\bibfnamefont
  {T.}~\bibnamefont {Horbury}}, \bibinfo {author} {\bibfnamefont
  {W.}~\bibnamefont {Matthaeus}}, \bibinfo {author} {\bibfnamefont
  {D.}~\bibnamefont {Burgess}}, \bibinfo {author} {\bibfnamefont
  {S.}~\bibnamefont {Servidio}},\ and\ \bibinfo {author} {\bibfnamefont
  {F.}~\bibnamefont {Valentini}},\ }\bibfield  {title} {\bibinfo {title} {On
  the transmission of turbulent structures across the earth's bow shock},\
  }\href {https://doi.org/10.3847/1538-4357/ac7798} {\bibfield  {journal}
  {\bibinfo  {journal} {The Astrophysical Journal}\ }\textbf {\bibinfo {volume}
  {933}},\ \bibinfo {pages} {167} (\bibinfo {year} {2022})}\BibitemShut
  {NoStop}%
\bibitem [{\citenamefont {{Ogilvie}}\ \emph {et~al.}(1995)\citenamefont
  {{Ogilvie}}, \citenamefont {{Chornay}}, \citenamefont {{Fritzenreiter}},
  \citenamefont {{Hunsaker}}, \citenamefont {{Keller}}, \citenamefont
  {{Lobell}}, \citenamefont {{Miller}}, \citenamefont {{Scudder}},
  \citenamefont {{Sittler}}, \citenamefont {{Torbert}}, \citenamefont
  {{Bodet}}, \citenamefont {{Needell}}, \citenamefont {{Lazarus}},
  \citenamefont {{Steinberg}}, \citenamefont {{Tappan}}, \citenamefont
  {{Mavretic}},\ and\ \citenamefont {{Gergin}}}]{ogilvie1995swe}%
  \BibitemOpen
  \bibfield  {author} {\bibinfo {author} {\bibfnamefont {K.~W.}\ \bibnamefont
  {{Ogilvie}}}, \bibinfo {author} {\bibfnamefont {D.~J.}\ \bibnamefont
  {{Chornay}}}, \bibinfo {author} {\bibfnamefont {R.~J.}\ \bibnamefont
  {{Fritzenreiter}}}, \bibinfo {author} {\bibfnamefont {F.}~\bibnamefont
  {{Hunsaker}}}, \bibinfo {author} {\bibfnamefont {J.}~\bibnamefont
  {{Keller}}}, \bibinfo {author} {\bibfnamefont {J.}~\bibnamefont {{Lobell}}},
  \bibinfo {author} {\bibfnamefont {G.}~\bibnamefont {{Miller}}}, \bibinfo
  {author} {\bibfnamefont {J.~D.}\ \bibnamefont {{Scudder}}}, \bibinfo {author}
  {\bibfnamefont {E.~C.}\ \bibnamefont {{Sittler}}, \bibfnamefont {Jr.}},
  \bibinfo {author} {\bibfnamefont {R.~B.}\ \bibnamefont {{Torbert}}}, \bibinfo
  {author} {\bibfnamefont {D.}~\bibnamefont {{Bodet}}}, \bibinfo {author}
  {\bibfnamefont {G.}~\bibnamefont {{Needell}}}, \bibinfo {author}
  {\bibfnamefont {A.~J.}\ \bibnamefont {{Lazarus}}}, \bibinfo {author}
  {\bibfnamefont {J.~T.}\ \bibnamefont {{Steinberg}}}, \bibinfo {author}
  {\bibfnamefont {J.~H.}\ \bibnamefont {{Tappan}}}, \bibinfo {author}
  {\bibfnamefont {A.}~\bibnamefont {{Mavretic}}},\ and\ \bibinfo {author}
  {\bibfnamefont {E.}~\bibnamefont {{Gergin}}},\ }\bibfield  {title} {\bibinfo
  {title} {{SWE, A Comprehensive Plasma Instrument for the Wind Spacecraft}},\
  }\href {https://doi.org/10.1007/BF00751326} {\bibfield  {journal} {\bibinfo
  {journal} {Space Science Reviews}\ }\textbf {\bibinfo {volume} {71}},\
  \bibinfo {pages} {55} (\bibinfo {year} {1995})}\BibitemShut {NoStop}%
\bibitem [{\citenamefont {{Lepping}}\ \emph {et~al.}(1995)\citenamefont
  {{Lepping}}, \citenamefont {{Ac{\~u}na}}, \citenamefont {{Burlaga}},
  \citenamefont {{Farrell}}, \citenamefont {{Slavin}}, \citenamefont
  {{Schatten}}, \citenamefont {{Mariani}}, \citenamefont {{Ness}},
  \citenamefont {{Neubauer}}, \citenamefont {{Whang}}, \citenamefont
  {{Byrnes}}, \citenamefont {{Kennon}}, \citenamefont {{Panetta}},
  \citenamefont {{Scheifele}},\ and\ \citenamefont
  {{Worley}}}]{lepping1995wind}%
  \BibitemOpen
  \bibfield  {author} {\bibinfo {author} {\bibfnamefont {R.~P.}\ \bibnamefont
  {{Lepping}}}, \bibinfo {author} {\bibfnamefont {M.~H.}\ \bibnamefont
  {{Ac{\~u}na}}}, \bibinfo {author} {\bibfnamefont {L.~F.}\ \bibnamefont
  {{Burlaga}}}, \bibinfo {author} {\bibfnamefont {W.~M.}\ \bibnamefont
  {{Farrell}}}, \bibinfo {author} {\bibfnamefont {J.~A.}\ \bibnamefont
  {{Slavin}}}, \bibinfo {author} {\bibfnamefont {K.~H.}\ \bibnamefont
  {{Schatten}}}, \bibinfo {author} {\bibfnamefont {F.}~\bibnamefont
  {{Mariani}}}, \bibinfo {author} {\bibfnamefont {N.~F.}\ \bibnamefont
  {{Ness}}}, \bibinfo {author} {\bibfnamefont {F.~M.}\ \bibnamefont
  {{Neubauer}}}, \bibinfo {author} {\bibfnamefont {Y.~C.}\ \bibnamefont
  {{Whang}}}, \bibinfo {author} {\bibfnamefont {J.~B.}\ \bibnamefont
  {{Byrnes}}}, \bibinfo {author} {\bibfnamefont {R.~S.}\ \bibnamefont
  {{Kennon}}}, \bibinfo {author} {\bibfnamefont {P.~V.}\ \bibnamefont
  {{Panetta}}}, \bibinfo {author} {\bibfnamefont {J.}~\bibnamefont
  {{Scheifele}}},\ and\ \bibinfo {author} {\bibfnamefont {E.~M.}\ \bibnamefont
  {{Worley}}},\ }\bibfield  {title} {\bibinfo {title} {{The Wind Magnetic Field
  Investigation}},\ }\href {https://doi.org/10.1007/BF00751330} {\bibfield
  {journal} {\bibinfo  {journal} {Space Science Reviews}\ }\textbf {\bibinfo
  {volume} {71}},\ \bibinfo {pages} {207} (\bibinfo {year} {1995})}\BibitemShut
  {NoStop}%
\bibitem [{\citenamefont {Jokipii}(1973)}]{jokipii1973turbulence}%
  \BibitemOpen
  \bibfield  {author} {\bibinfo {author} {\bibfnamefont {J.~R.}\ \bibnamefont
  {Jokipii}},\ }\bibfield  {title} {\bibinfo {title} {Turbulence and
  scintillations in the interplanetary plasma},\ }\href
  {https://doi.org/10.1146/annurev.aa.11.090173.000245} {\bibfield  {journal}
  {\bibinfo  {journal} {Annual Review of Astronomy and Astrophysics}\ }\textbf
  {\bibinfo {volume} {11}},\ \bibinfo {pages} {1} (\bibinfo {year}
  {1973})}\BibitemShut {NoStop}%
\bibitem [{\citenamefont {Matthaeus}\ \emph {et~al.}(1999)\citenamefont
  {Matthaeus}, \citenamefont {Smith},\ and\ \citenamefont
  {Bieber}}]{matthaeus1999correlation}%
  \BibitemOpen
  \bibfield  {author} {\bibinfo {author} {\bibfnamefont {W.~H.}\ \bibnamefont
  {Matthaeus}}, \bibinfo {author} {\bibfnamefont {C.~W.}\ \bibnamefont
  {Smith}},\ and\ \bibinfo {author} {\bibfnamefont {J.~W.}\ \bibnamefont
  {Bieber}},\ }\bibfield  {title} {\bibinfo {title} {Correlation lengths, the
  ultrascale, and the spatial structure of interplanetary turbulence},\ }\href
  {https://doi.org/10.1063/1.58686} {\bibfield  {journal} {\bibinfo  {journal}
  {AIP Conference Proceedings}\ }\textbf {\bibinfo {volume} {471}},\ \bibinfo
  {pages} {511} (\bibinfo {year} {1999})}\BibitemShut {NoStop}%
\bibitem [{\citenamefont {Servidio}\ \emph
  {et~al.}(2010{\natexlab{b}})\citenamefont {Servidio}, \citenamefont
  {Matthaeus}, \citenamefont {Shay}, \citenamefont {Dmitruk}, \citenamefont
  {Cassak},\ and\ \citenamefont {Wan}}]{servidio2010statistics}%
  \BibitemOpen
  \bibfield  {author} {\bibinfo {author} {\bibfnamefont {S.}~\bibnamefont
  {Servidio}}, \bibinfo {author} {\bibfnamefont {W.~H.}\ \bibnamefont
  {Matthaeus}}, \bibinfo {author} {\bibfnamefont {M.~A.}\ \bibnamefont {Shay}},
  \bibinfo {author} {\bibfnamefont {P.}~\bibnamefont {Dmitruk}}, \bibinfo
  {author} {\bibfnamefont {P.~A.}\ \bibnamefont {Cassak}},\ and\ \bibinfo
  {author} {\bibfnamefont {M.}~\bibnamefont {Wan}},\ }\bibfield  {title}
  {\bibinfo {title} {Statistics of magnetic reconnection in two-dimensional
  magnetohydrodynamic turbulence},\ }\href {https://doi.org/10.1063/1.3368798}
  {\bibfield  {journal} {\bibinfo  {journal} {Physics of Plasmas}\ }\textbf
  {\bibinfo {volume} {17}},\ \bibinfo {eid} {032315} (\bibinfo {year}
  {2010}{\natexlab{b}})}\BibitemShut {NoStop}%
\bibitem [{\citenamefont {Pecora}\ \emph {et~al.}(2019)\citenamefont {Pecora},
  \citenamefont {Greco}, \citenamefont {Hu}, \citenamefont {Servidio},
  \citenamefont {Chasapis},\ and\ \citenamefont
  {Matthaeus}}]{pecora2019single}%
  \BibitemOpen
  \bibfield  {author} {\bibinfo {author} {\bibfnamefont {F.}~\bibnamefont
  {Pecora}}, \bibinfo {author} {\bibfnamefont {A.}~\bibnamefont {Greco}},
  \bibinfo {author} {\bibfnamefont {Q.}~\bibnamefont {Hu}}, \bibinfo {author}
  {\bibfnamefont {S.}~\bibnamefont {Servidio}}, \bibinfo {author}
  {\bibfnamefont {A.~G.}\ \bibnamefont {Chasapis}},\ and\ \bibinfo {author}
  {\bibfnamefont {W.~H.}\ \bibnamefont {Matthaeus}},\ }\bibfield  {title}
  {\bibinfo {title} {Single-spacecraft identification of flux tubes and current
  sheets in the solar wind},\ }\href
  {https://doi.org/doi.org/10.3847/2041-8213/ab32d9} {\bibfield  {journal}
  {\bibinfo  {journal} {The Astrophysical Journal Letters}\ }\textbf {\bibinfo
  {volume} {881}},\ \bibinfo {pages} {L11} (\bibinfo {year}
  {2019})}\BibitemShut {NoStop}%
\bibitem [{\citenamefont {Yang}\ \emph {et~al.}(2017)\citenamefont {Yang},
  \citenamefont {Matthaeus}, \citenamefont {Parashar}, \citenamefont
  {Haggerty}, \citenamefont {Roytershteyn}, \citenamefont {Daughton},
  \citenamefont {Wan}, \citenamefont {Shi},\ and\ \citenamefont
  {Chen}}]{yang2017energy}%
  \BibitemOpen
  \bibfield  {author} {\bibinfo {author} {\bibfnamefont {Y.}~\bibnamefont
  {Yang}}, \bibinfo {author} {\bibfnamefont {W.~H.}\ \bibnamefont {Matthaeus}},
  \bibinfo {author} {\bibfnamefont {T.~N.}\ \bibnamefont {Parashar}}, \bibinfo
  {author} {\bibfnamefont {C.~C.}\ \bibnamefont {Haggerty}}, \bibinfo {author}
  {\bibfnamefont {V.}~\bibnamefont {Roytershteyn}}, \bibinfo {author}
  {\bibfnamefont {W.}~\bibnamefont {Daughton}}, \bibinfo {author}
  {\bibfnamefont {M.}~\bibnamefont {Wan}}, \bibinfo {author} {\bibfnamefont
  {Y.}~\bibnamefont {Shi}},\ and\ \bibinfo {author} {\bibfnamefont
  {S.}~\bibnamefont {Chen}},\ }\bibfield  {title} {\bibinfo {title} {Energy
  transfer, pressure tensor, and heating of kinetic plasma},\ }\href
  {https://doi.org/10.1063/1.4990421} {\bibfield  {journal} {\bibinfo
  {journal} {Physics of Plasmas}\ }\textbf {\bibinfo {volume} {24}},\ \bibinfo
  {pages} {072306} (\bibinfo {year} {2017})}\BibitemShut {NoStop}%
\bibitem [{\citenamefont {Bandyopadhyay}\ \emph
  {et~al.}(2020{\natexlab{a}})\citenamefont {Bandyopadhyay}, \citenamefont
  {Matthaeus}, \citenamefont {Parashar}, \citenamefont {Yang}, \citenamefont
  {Chasapis}, \citenamefont {Giles}, \citenamefont {Gershman}, \citenamefont
  {Pollock}, \citenamefont {Russell}, \citenamefont {Strangeway}, \citenamefont
  {Torbert}, \citenamefont {Moore},\ and\ \citenamefont
  {Burch}}]{bandyopadhyay2020statistics}%
  \BibitemOpen
  \bibfield  {author} {\bibinfo {author} {\bibfnamefont {R.}~\bibnamefont
  {Bandyopadhyay}}, \bibinfo {author} {\bibfnamefont {W.~H.}\ \bibnamefont
  {Matthaeus}}, \bibinfo {author} {\bibfnamefont {T.~N.}\ \bibnamefont
  {Parashar}}, \bibinfo {author} {\bibfnamefont {Y.}~\bibnamefont {Yang}},
  \bibinfo {author} {\bibfnamefont {A.}~\bibnamefont {Chasapis}}, \bibinfo
  {author} {\bibfnamefont {B.~L.}\ \bibnamefont {Giles}}, \bibinfo {author}
  {\bibfnamefont {D.~J.}\ \bibnamefont {Gershman}}, \bibinfo {author}
  {\bibfnamefont {C.~J.}\ \bibnamefont {Pollock}}, \bibinfo {author}
  {\bibfnamefont {C.~T.}\ \bibnamefont {Russell}}, \bibinfo {author}
  {\bibfnamefont {R.~J.}\ \bibnamefont {Strangeway}}, \bibinfo {author}
  {\bibfnamefont {R.~B.}\ \bibnamefont {Torbert}}, \bibinfo {author}
  {\bibfnamefont {T.~E.}\ \bibnamefont {Moore}},\ and\ \bibinfo {author}
  {\bibfnamefont {J.~L.}\ \bibnamefont {Burch}},\ }\bibfield  {title} {\bibinfo
  {title} {Statistics of kinetic dissipation in the earth's magnetosheath: Mms
  observations},\ }\href {https://doi.org/10.1103/PhysRevLett.124.255101}
  {\bibfield  {journal} {\bibinfo  {journal} {Phys. Rev. Lett.}\ }\textbf
  {\bibinfo {volume} {124}},\ \bibinfo {pages} {255101} (\bibinfo {year}
  {2020}{\natexlab{a}})}\BibitemShut {NoStop}%
\bibitem [{\citenamefont {{Yang}}\ \emph {et~al.}(2017)\citenamefont {{Yang}},
  \citenamefont {{Matthaeus}}, \citenamefont {{Parashar}}, \citenamefont
  {{Wu}}, \citenamefont {{Wan}}, \citenamefont {{Shi}}, \citenamefont {{Chen}},
  \citenamefont {{Roytershteyn}},\ and\ \citenamefont
  {{Daughton}}}]{YangEA17-PRE}%
  \BibitemOpen
  \bibfield  {author} {\bibinfo {author} {\bibfnamefont {Y.}~\bibnamefont
  {{Yang}}}, \bibinfo {author} {\bibfnamefont {W.~H.}\ \bibnamefont
  {{Matthaeus}}}, \bibinfo {author} {\bibfnamefont {T.~N.}\ \bibnamefont
  {{Parashar}}}, \bibinfo {author} {\bibfnamefont {P.}~\bibnamefont {{Wu}}},
  \bibinfo {author} {\bibfnamefont {M.}~\bibnamefont {{Wan}}}, \bibinfo
  {author} {\bibfnamefont {Y.}~\bibnamefont {{Shi}}}, \bibinfo {author}
  {\bibfnamefont {S.}~\bibnamefont {{Chen}}}, \bibinfo {author} {\bibfnamefont
  {V.}~\bibnamefont {{Roytershteyn}}},\ and\ \bibinfo {author} {\bibfnamefont
  {W.}~\bibnamefont {{Daughton}}},\ }\bibfield  {title} {\bibinfo {title}
  {{Energy transfer channels and turbulence cascade in Vlasov-Maxwell
  turbulence}},\ }\href {https://doi.org/10.1103/PhysRevE.95.061201} {\bibfield
   {journal} {\bibinfo  {journal} {Phys. Rev. E}\ }\textbf {\bibinfo {volume}
  {95}},\ \bibinfo {eid} {061201} (\bibinfo {year} {2017})}\BibitemShut
  {NoStop}%
\bibitem [{\citenamefont {Bandyopadhyay}\ \emph
  {et~al.}(2020{\natexlab{b}})\citenamefont {Bandyopadhyay}, \citenamefont
  {Matthaeus}, \citenamefont {Parashar}, \citenamefont {Yang}, \citenamefont
  {Chasapis}, \citenamefont {Giles}, \citenamefont {Gershman}, \citenamefont
  {Pollock}, \citenamefont {Russell}, \citenamefont {Strangeway}, \citenamefont
  {Torbert}, \citenamefont {Moore},\ and\ \citenamefont
  {Burch}}]{BandyopadhyayEA20-PiD}%
  \BibitemOpen
  \bibfield  {author} {\bibinfo {author} {\bibfnamefont {R.}~\bibnamefont
  {Bandyopadhyay}}, \bibinfo {author} {\bibfnamefont {W.~H.}\ \bibnamefont
  {Matthaeus}}, \bibinfo {author} {\bibfnamefont {T.~N.}\ \bibnamefont
  {Parashar}}, \bibinfo {author} {\bibfnamefont {Y.}~\bibnamefont {Yang}},
  \bibinfo {author} {\bibfnamefont {A.}~\bibnamefont {Chasapis}}, \bibinfo
  {author} {\bibfnamefont {B.~L.}\ \bibnamefont {Giles}}, \bibinfo {author}
  {\bibfnamefont {D.~J.}\ \bibnamefont {Gershman}}, \bibinfo {author}
  {\bibfnamefont {C.~J.}\ \bibnamefont {Pollock}}, \bibinfo {author}
  {\bibfnamefont {C.~T.}\ \bibnamefont {Russell}}, \bibinfo {author}
  {\bibfnamefont {R.~J.}\ \bibnamefont {Strangeway}}, \bibinfo {author}
  {\bibfnamefont {R.~B.}\ \bibnamefont {Torbert}}, \bibinfo {author}
  {\bibfnamefont {T.~E.}\ \bibnamefont {Moore}},\ and\ \bibinfo {author}
  {\bibfnamefont {J.~L.}\ \bibnamefont {Burch}},\ }\bibfield  {title} {\bibinfo
  {title} {Statistics of kinetic dissipation in the earth's magnetosheath: Mms
  observations},\ }\href {https://doi.org/10.1103/PhysRevLett.124.255101}
  {\bibfield  {journal} {\bibinfo  {journal} {Phys. Rev. Lett.}\ }\textbf
  {\bibinfo {volume} {124}},\ \bibinfo {pages} {255101} (\bibinfo {year}
  {2020}{\natexlab{b}})}\BibitemShut {NoStop}%
\bibitem [{\citenamefont {Stawarz}\ \emph {et~al.}(2022)\citenamefont
  {Stawarz}, \citenamefont {Eastwood}, \citenamefont {Phan}, \citenamefont
  {Gingell}, \citenamefont {Pyakurel}, \citenamefont {Shay}, \citenamefont
  {Robertson}, \citenamefont {Russell},\ and\ \citenamefont
  {Le~Contel}}]{StawarzEA22}%
  \BibitemOpen
  \bibfield  {author} {\bibinfo {author} {\bibfnamefont {J.~E.}\ \bibnamefont
  {Stawarz}}, \bibinfo {author} {\bibfnamefont {J.~P.}\ \bibnamefont
  {Eastwood}}, \bibinfo {author} {\bibfnamefont {T.~D.}\ \bibnamefont {Phan}},
  \bibinfo {author} {\bibfnamefont {I.~L.}\ \bibnamefont {Gingell}}, \bibinfo
  {author} {\bibfnamefont {P.~S.}\ \bibnamefont {Pyakurel}}, \bibinfo {author}
  {\bibfnamefont {M.~A.}\ \bibnamefont {Shay}}, \bibinfo {author}
  {\bibfnamefont {S.~L.}\ \bibnamefont {Robertson}}, \bibinfo {author}
  {\bibfnamefont {C.~T.}\ \bibnamefont {Russell}},\ and\ \bibinfo {author}
  {\bibfnamefont {O.}~\bibnamefont {Le~Contel}},\ }\bibfield  {title} {\bibinfo
  {title} {Turbulence-driven magnetic reconnection and the magnetic correlation
  length: Observations from magnetospheric multiscale in earth's
  magnetosheath},\ }\href {https://doi.org/10.1063/5.0071106} {\bibfield
  {journal} {\bibinfo  {journal} {Physics of Plasmas}\ }\textbf {\bibinfo
  {volume} {29}},\ \bibinfo {pages} {012302} (\bibinfo {year} {2022})},\
  \Eprint {https://arxiv.org/abs/https://doi.org/10.1063/5.0071106}
  {https://doi.org/10.1063/5.0071106} \BibitemShut {NoStop}%
\bibitem [{\citenamefont {Kerr}(1985)}]{kerr1985higher}%
  \BibitemOpen
  \bibfield  {author} {\bibinfo {author} {\bibfnamefont {R.~M.}\ \bibnamefont
  {Kerr}},\ }\bibfield  {title} {\bibinfo {title} {Higher-order derivative
  correlations and the alignment of small-scale structures in isotropic
  numerical turbulence},\ }\href {https://doi.org/10.1017/S0022112085001136}
  {\bibfield  {journal} {\bibinfo  {journal} {Journal of Fluid Mechanics}\
  }\textbf {\bibinfo {volume} {153}},\ \bibinfo {pages} {31–58} (\bibinfo
  {year} {1985})}\BibitemShut {NoStop}%
\bibitem [{\citenamefont {Blackburn}\ \emph {et~al.}(1996)\citenamefont
  {Blackburn}, \citenamefont {Mansour},\ and\ \citenamefont
  {Cantwell}}]{blackburn1996topology}%
  \BibitemOpen
  \bibfield  {author} {\bibinfo {author} {\bibfnamefont {H.~M.}\ \bibnamefont
  {Blackburn}}, \bibinfo {author} {\bibfnamefont {N.~N.}\ \bibnamefont
  {Mansour}},\ and\ \bibinfo {author} {\bibfnamefont {B.~J.}\ \bibnamefont
  {Cantwell}},\ }\bibfield  {title} {\bibinfo {title} {Topology of fine-scale
  motions in turbulent channel flow},\ }\href
  {https://doi.org/10.1017/S0022112096001802} {\bibfield  {journal} {\bibinfo
  {journal} {Journal of Fluid Mechanics}\ }\textbf {\bibinfo {volume} {310}},\
  \bibinfo {pages} {269–292} (\bibinfo {year} {1996})}\BibitemShut {NoStop}%
\bibitem [{\citenamefont {Consolini}\ \emph {et~al.}(2015)\citenamefont
  {Consolini}, \citenamefont {Materassi}, \citenamefont {Marcucci},\ and\
  \citenamefont {Pallocchia}}]{consolini2015statistics}%
  \BibitemOpen
  \bibfield  {author} {\bibinfo {author} {\bibfnamefont {G.}~\bibnamefont
  {Consolini}}, \bibinfo {author} {\bibfnamefont {M.}~\bibnamefont
  {Materassi}}, \bibinfo {author} {\bibfnamefont {M.~F.}\ \bibnamefont
  {Marcucci}},\ and\ \bibinfo {author} {\bibfnamefont {G.}~\bibnamefont
  {Pallocchia}},\ }\bibfield  {title} {\bibinfo {title} {Statistics of the
  velocity gradient tensor in space plasma turbulent flows},\ }\href
  {https://doi.org/10.1088/0004-637X/812/1/84} {\bibfield  {journal} {\bibinfo
  {journal} {The Astrophysical Journal}\ }\textbf {\bibinfo {volume} {812}},\
  \bibinfo {pages} {84} (\bibinfo {year} {2015})}\BibitemShut {NoStop}%
\bibitem [{\citenamefont {{Spence}}(2019)}]{SpenceEA19}%
  \BibitemOpen
  \bibfield  {author} {\bibinfo {author} {\bibfnamefont {H.~E.}\ \bibnamefont
  {{Spence}}},\ }\bibfield  {title} {\bibinfo {title} {{HelioSwarm: Unlocking
  the Multiscale Mysteries of Weakly-Collisional Magnetized Plasma Turbulence
  and Ion Heating}},\ }in\ \href@noop {} {\emph {\bibinfo {booktitle} {AGU Fall
  Meeting Abstracts}}},\ Vol.\ \bibinfo {volume} {2019}\ (\bibinfo {year}
  {2019})\ pp.\ \bibinfo {pages} {SH11B--04}\BibitemShut {NoStop}%
\end{thebibliography}

 \newcommand{\BIBand} {and} 
  \newcommand{\boldVol}[1] {\textbf{#1}} 
  \providecommand{\SortNoop}[1]{} 
  \providecommand{\sortnoop}[1]{} 
  \newcommand{\stereo} {\emph{{S}{T}{E}{R}{E}{O}}} 
  \newcommand{\au} {{A}{U}\ } 
  \newcommand{\AU} {{A}{U}\ } 
  \newcommand{\MHD} {{M}{H}{D}\ } 
  \newcommand{\mhd} {{M}{H}{D}\ } 
  \newcommand{\RMHD} {{R}{M}{H}{D}\ } 
  \newcommand{\rmhd} {{R}{M}{H}{D}\ } 
  \newcommand{\wkb} {{W}{K}{B}\ } 
  \newcommand{\alfven} {{A}lfv{\'e}n\ } 
  \newcommand{\alfvenic} {{A}lfv{\'e}nic\ } 
  \newcommand{\Alfven} {{A}lfv{\'e}n\ } 
  \newcommand{\Alfvenic} {{A}lfv{\'e}nic\ }

\end{document}